\documentclass[%
reprint,
superscriptaddress,
notitlepage,
nofootinbib,
 amsmath,amssymb,amsfonts
floatfix,
]{revtex4-1}
\usepackage{mathtools}
\usepackage{amsthm}
\usepackage{appendix}
\usepackage{physics}
\usepackage{graphicx}
\usepackage{dcolumn}
\usepackage{bm}
\usepackage{hyperref}
\usepackage[normalem]{ulem}
\usepackage{xcolor}

\newtheorem{rem}{Remark}

\begin{document}

\preprint{}

\title{Exact steady state of the open $XX$-spin chain:\\ entanglement and transport properties}

\author{F. Benatti}
\affiliation{Department of Physics, University of Trieste, I-34151 Trieste, Italy}
\affiliation{Istituto Nazionale di Fisica Nucleare (INFN), Sezione di Trieste, I-34151 Trieste, Italy}
\author{R. Floreanini}
\affiliation{Istituto Nazionale di Fisica Nucleare (INFN), Sezione di Trieste, I-34151 Trieste, Italy}
\email{f@infn}
\author{L. Memarzadeh}%
\email{memarzadeh@sharif.edu\\The authors contributed equally to this study and are listed in
alphabetical order.}
\affiliation{Department of Physics, Sharif University of Technology, Tehran, Iran}


\begin{abstract}
We study the reduced dynamics of open quantum spin chains of arbitrary length $N$ with nearest neighbour $XX$ interactions, immersed within an external constant magnetic field along the $z$ direction, whose end spins are weakly coupled to heat baths at different temperatures, via energy preserving couplings.
We find the analytic expression of the unique  stationary state of the master equation obtained  in the so-called global approach based on the spectralization of the full chain Hamiltonian.
Hinging upon the explicit stationary state, we reveal the presence of sink and source terms in the spin-flow continuity equation and compare their behaviour with that of the stationary heat flow.
Moreover, we also obtain analytic expressions for the steady state two-spin reduced density matrices and for their concurrence. We then set up an algorithm suited to compute the stationary  bipartite entanglement along the chain and to study its dependence on the Hamiltonian parameters and on the bath temperatures.
\end{abstract}

\maketitle


\section{Introduction}
\label{sec:introduction}
Transport phenomena in open interacting quantum spin chains have recently received an increasing attention as  instances of many-body systems driven by intrinsic
inter-spin interactions and coupled 
to external heat baths at the two ends of the chain.
Specific experimental realizations have been reported in scenarios involving ultracold-atoms, light-harvesting complexes and quantum thermodynamics at large~\cite{Datta}-\cite{Prosen5}.

In presence of external baths, the reduced dynamics of any open quantum system is obtained by tracing over the baths' degrees of freedom. When the strength of the system-baths interaction is small, applying the weak-coupling limit techniques yields a dissipative irreversible time evolution that is generated by a master equation in Gorini-Kossakowski-Sudarshan-Lindblad (GKSL) form~\cite{Alicki-Lendi}-\cite{Merkli}.

The derivation of the GKSL master equation requires the diagonalization of the full spin-chain Hamiltonian. 
Due to the many degrees of freedom of the quantum spin chain and their mutual interactions, 
dissipative effects then arise involving all spins in the chain  together with environment-induced excitation transfer between different sites ({\it e.g.} see \cite{Davies4}-\cite{Rivas2}). These gives rise to new, global effects in transport phenomena
that can not be captured using other, simplified approaches to the chain open dynamics.%
\footnote{
As finding eigenvalues and eigenvectors of the system Hamiltonian might in general be laboriously difficult, 
an alternative approach has been often advocated, consisting in neglecting the inter-spin interaction in the derivation
of the master equa-tion ({\it e.g.} see \cite{Michel}-\cite{Hovhannisyan}). Although the two approaches, named global and local, are regularly adopted in applications and compared~\cite{Rivas1,Guimaraes,Werlang,Santos,Migliore,Zoubi}, \cite{Levy}-\cite{Giovannetti20},
it turns out that the local approach might not be able to capture all the correct system transport properties~\cite{BFM}.
}

In the following, we focus on the study of the stationary transport and bipartite entanglement properties of open $XX$-chains with energy conserving couplings to external baths. We provide an explicit analytic form for the chain stationary state by means of which we obtain analytic expressions for the spin flow, revealing the presence of sink and source terms, and for the heat flow.
Remarkably, we have also been able to explicitly compute the reduced two-spin density matrices resulting from the stationary state and study the corresponding bipartite entanglement along the chain. For the latter task, we develop a suitable algorithmic representation of the stationary state in the spin representation.\\
\indent
The structure of the paper is as follows: in section II we set the framework for the derivation of the open chain dynamics and diagonalize the chain Hamiltonian by turning the spin representation into a Fermionic one. In Section III we derive the Lindblad operators yielding the dissipative contribution to the GKSL master equation, prove that the latter has a unique stationary state and explicitly derive its expression in the Fermionic representation. Based on it, we then discuss the ensuing transport properties in terms of spin and heat flows.
In section IV we rewrite the stationary state in the spin representation and show that
it provides reduced two-spin density matrices in the so-called $X$ form which allows for a simple analytical expression of the concurrence. Then, we set up a representation of the stationary state that is best suited for the study of bipartite entanglement and its dependence on the various parameters of the chain and on the temperatures of the baths coupled to it.
We conclude by summarizing and discussing the results, while the more technical issues are presented in various Appendices.

\section{Open $XX$ spin chain of length $N$}
\label{sec:system}
As mentioned in the Introduction, in the following we address an 
open quantum chain consisting of $N$ spins at sites $1,2,\ldots, N$, immersed in a constant magnetic field along the $z$ direction, with $XX$ nearest neighbour interactions among themselves. 
The ensuing closed chain dynamics is thus generated by the following Hamiltonian:
\begin{equation}
H=g\sum_{\ell=1}^{N-1}\left(\sigma_{x}^{(\ell)}\sigma_{x}^{(\ell+1)}+\sigma_{y}^{(\ell)}\sigma_{y}^{(\ell+1)}\right)+
\Delta\sum_{\ell=1}^N\sigma_z^{(\ell)}\ ,
\label{spin-hamiltonian}
\end{equation}
with free boundary ocnditions, where $\Delta> 0$ is the intensity of the constant transverse magnetic field, $\sigma_{x,y,z}^{(\ell)}$ are the Pauli matrices at site $\ell$, and $g>0$ is the strength of the nearest neighbour interaction.
Throughout the paper we work in natural units where both Planck and Boltzmann constants are set to 1, $\hbar=\kappa_B=1$.

The spin chain is then turned into an open many-body quantum system by coupling the two end spins, at site~1 on the left end, $L$, and at site $N$ on the right end, $R$, to two independent free Bosonic thermal baths with Hamiltonians 
\begin{equation}
\label{freeHam}
H_\alpha=\int_{0}^{+\infty}{\rm d}\nu\,\nu\, \mathfrak{b}^\dag_\alpha(\nu)\,\mathfrak{b}_\alpha(\nu)\ ,\qquad \alpha=L,R\ ,
\end{equation}
where $\mathfrak{b}_\alpha(\nu)$, $\mathfrak{b}^\dagger_\alpha(\nu)$ 
are Bosonic operators satisfying the canonical commutation relations
$$
[\mathfrak{b}_\alpha(\nu),\, \mathfrak{b}_\beta^\dag(\nu')]=\delta_{\alpha\beta}\, \delta(\nu-\nu') \ .
$$
The coupling of the baths to the left and right spins are described by the interaction Hamiltonian
\begin{equation}
H'=\lambda\,\sum_{\alpha=L,R} \Big( \sigma_+^{(\alpha)} \mathfrak{B}_\alpha + \sigma_-^{(\alpha)} \mathfrak{B}_\alpha^\dag \Big)\ ,
\label{interaction-h}
\end{equation}
where $\lambda<<1$ is a dimensionless coupling constant,
\begin{equation}
\sigma_\pm^{(\ell)} \equiv \frac{1}{2}\big( \sigma^{(\ell)}_x\pm i \sigma^{(\ell)} _y\big)\ ,
\label{sigmas}
\end{equation}
are spin ladder operators at site $\ell$, whence $\sigma_\pm^{(L)}=\sigma_\pm^{(1)}$ and $\sigma_\pm^{(R)}=\sigma_\pm^{(N)}$, while
\begin{equation}
\mathfrak{B}_\alpha=\int_0^\infty {\rm d}\nu\, h_\alpha(\nu)\, \mathfrak{b}_\alpha(\nu)\ ,\quad [h_\alpha(\nu)]^* = h_\alpha(\nu)\ ,
\label{bath-operator}
\end{equation}
are bath operators, with $*$ meaning complex conjugation and $h_{L,R}(\nu)$ are suitable smearing functions.
%
%
Referring to~\cite{BFM} for more details, we begin by shortly reviewing the rigorous weak-coupling limit derivation of  the open chain master equation of GKSL type in the so-called global approach.
As we shall show, the resulting dissipative dynamics of the $N$ spins in the presence of the two baths involves the full inter-spin interactions.

Assuming the free Boson baths to be  in their equilibrium Gibbs states at temperatures $T_L\equiv1/\beta_L$ and
$T_R\equiv1/\beta_R$, the state of the environment is then given by 
\begin{equation}
\label{Gibbs}
\rho_{env}=\frac{{\rm e}^{-\beta_L\,H_L}}{{\rm Tr}\Big({\rm e}^{-\beta_L\,H_L}\Big)}\,\otimes\,
\frac{{\rm e}^{-\beta_R\,H_R}}{{\rm Tr}\Big({\rm e}^{-\beta_R\,H_R}\Big)}\ .
\end{equation}
It  is invariant under the bath dynamics generated by $H_{env}=\sum_{\alpha=L,R}H_\alpha$
and exhibits thermal expectations of the form
\begin{align}
\label{expectation1}
&\hskip-.2cm
{\rm Tr}_B\Big(\rho_{env}b^\dag_\alpha(\nu)b_{\alpha'}(\nu')\Big)=\delta_{\alpha\alpha'}\delta(\nu-\nu')\,n_\alpha(\nu)\\
\label{expectation2}
&\hskip-.2cm
{\rm Tr}_B\Big(\rho_{env}b_\alpha(\nu)b^\dag_{\alpha'}(\nu')\Big)=\delta_{\alpha\alpha'}\delta(\nu-\nu')\,(1+n_\alpha(\nu)),
\end{align}
with thermal mean occupation numbers
\begin{equation}
\label{thermnumb}
n_\alpha(\nu)=\frac{1}{{\rm e}^{\beta_\alpha\nu}-1}\ ,\qquad\nu\geq 0\ .
\end{equation}
Finally, choosing the initial state of the compound system chain plus baths of the form $\rho_{\rm tot}(0)=\rho(0)\otimes \rho_{env}$, with $\rho(0)$ an initial state of the $N$ spins of the chain, in presence of a fast decay of the thermal correlation functions, one applies the weak-coupling limit techniques and obtains a fully physically consistent dissipative chain dynamics~\cite{Alicki-Lendi}-\cite{Merkli}.
In practice, the initial state of the compound system spin-chain plus baths evolves into 
\begin{equation}
\label{toev}
\rho_{\rm tot}(t)={\rm e}^{-itH_{\rm tot}}\,\rho_{\rm tot}(0)\,{\rm e}^{itH_{\rm tot}}\ ,
\end{equation}
where $H_{\rm tot}=H+H_{env}+H'$ is the total system Hamiltonian. The state of the open chain at time $t$, $\rho(t)$, is then 
retrieved by tracing over the baths degrees of freedom, $\rho(t)={\rm Tr}_{env}\Big(\rho_{\rm tot}(t)\Big)$.
Then, one rescales the physical time variable $t$ to $\tau=t\,\lambda^2$ and takes the limit $\lambda\to 0$ in
$\partial_t{\rm Tr}_{env}\Big(\rho_{\rm tot}(t)\Big)$. In doing so, too fast oscillations with respect to the chain transition frequencies $\omega=E_i-E_j$ are suppressed, where $E_i$ and $\vert E_i\rangle$ solve the spin Hamiltonian eigenvalue equation $H\vert E_i\rangle=E_i\,\vert E_i\rangle$.
This procedure corresponds to  the so-called rotating wave approximation leading to
a master equation of the GKSL form 
\begin{equation}
\label{MS1}
{\partial\rho(t)\over \partial t}=-i \big[H+\lambda^2\,H_{LS},\rho(t)]+{\mathbb D}[\rho(t)]=\mathbb{L}[\rho(t)]\ .
\end{equation}
On the right hand side of the above time-evolution equation, one distinguishes a Hamiltonian term $\lambda^2\,H_{LS}$ which provides 
a Lamb-shift correction to the spin-chain Hamiltonian $H$  and a purely dissipative term ${\mathbb D}[\rho(t)]$.
As we shall see, in the specific physical context here considered,  ${\mathbb D}[\rho(t)]$ consists of contributions ${\mathbb D}^{(\alpha)}_{\omega}[\rho(t)]$ 
resulting from positive transition frequencies $\omega\geq 0$, only:
\begin{equation}
{\mathbb D}[\rho(t)]=\lambda^2 \sum_{\alpha=L,R} \ \sum_{\omega\geq 0}{\mathbb D}^{(\alpha)}_{\omega}[\rho(t)]\ .
\label{MS2}
\end{equation}
%
%
Their explicit form reads
\begin{eqnarray}
\nonumber 
&&\hskip -1cm
{\mathbb D}^{(\alpha)}_{\omega}[\rho(t)]= C^{(\alpha)}_{\omega} \bigg[ A_\alpha(\omega) \rho(t) A_\alpha^\dag(\omega)\\ 
\label{MS3a}
&&\hskip 2cm
-\,\frac{1}{2}\bigg\{A_\alpha^\dag(\omega)A_\alpha(\omega), \rho(t)\bigg\}\bigg]\\
\nonumber
&&\hskip 1cm
+\widetilde C^{(\alpha)}_{\omega} \bigg[ A_\alpha^\dag(\omega) \rho(t) A_\alpha(\omega)\\
&&
\hskip 2cm
-\,\frac{1}{2}\bigg\{A_\alpha(\omega)A_\alpha^\dag(\omega), \rho(t)\bigg\}\bigg]\ ,
\label{MS3b}
\end{eqnarray}
whose coefficients 
\begin{eqnarray}
\label{MS4a}
C^{(\alpha)}_{\omega}&=& 2\pi\, [h_\alpha(\omega)]^2\,  \big(n_\alpha(\omega)+1\big)\ ,\\
\label{MS4b}
\widetilde C^{(\alpha)}_{\omega}&=&2\pi\, [h_\alpha(\omega)]^2\,  n_\alpha(\omega)\ ,
\end{eqnarray}
with  $n_\alpha(\omega)$ as in~\eqref{thermnumb}, come from the real parts of the half-Fourier transforms of the bath correlation functions.


Instead, the Lamb-shift correction amounts to the Hamiltonian
\begin{eqnarray}
\nonumber
&&
H_{LS} = \sum_{\alpha=L,R}\sum_{\omega}\ \bigg[S^{(\alpha)}_{\omega} A_\alpha^\dag(\omega)A_\alpha(\omega)\\
&&\hskip 2 cm
+\, \widetilde S^{(\alpha)}_{\omega} A_\alpha(\omega)A_\alpha^\dag(\omega)\bigg]\ ,
\label{MS5}
\end{eqnarray}
where, unlike the dissipative term, the sum runs now over all positive and negative transition frequencies and whose coefficients read 
\begin{eqnarray}
\label{MS6a}
S^{(\alpha)}_{\omega}&=& P\,\int_{0}^{+\infty} {\rm d}\nu\,[h_\alpha(\nu)]^2 \frac{1+n_\alpha(\nu)}{\omega-\nu}\ ,\\
\label{MS6b}
\widetilde S^{(\alpha)}_{\omega}&=& P\,\int_{0}^{+\infty} {\rm d}\nu\,[h_\alpha(\nu)]^2 \frac{n_\alpha(\nu)}{\nu-\omega}\ ,
\end{eqnarray}
with $P$ denoting the principal value.

In all the previous expressions there appear Lindblad operators of the form
\begin{equation}
\label{Lindop1a}
A^\dag_\alpha(\omega)=\sum_{E_i-E_j=\omega} \vert E_i\rangle\langle E_i\vert\,\sigma_+^{(\alpha)}\,
\vert E_j\rangle\langle E_j\vert\ ,
\end{equation}
together with their Hermitean conjugates
\begin{equation}
\label{Lindop1a}
A_\alpha(\omega)=\sum_{E_i-E_j=\omega} \vert E_j\rangle\langle E_j\vert\,\sigma_-^{(\alpha)}\,
\vert E_i\rangle\langle E_i\vert\ .
\end{equation}
In order to obtain explicit
expressions for the elements of the master equation (\ref{MS1}), one needs to work with eigenvalues and eigenvectors of the full spin Hamiltonian $H$: this point of view is known as
\textit{global approach} to open quantum spin chains. This way of proceeding is in contrast with  the so-called \textit{local approach} where the weak-coupling limit 
is implemented by switching off the spin interactions, thus obtaining strictly local dissipative terms that involve only the left and the right spins.
The spin interactions are then reinserted at the end of the weak-coupling procedure.

\begin{rem}
\label{remint}
The fact that the dissipative contribution to the generator, ${\mathbb D}[\rho(t)]$,  involves only transition frequencies $\omega\geq 0$ is due the thermal bath energies  being 
positive and to the form of the interaction in~\eqref{interaction-h}. Indeed, in the interaction representation, terms  as $A^\dag_\alpha(\omega)\mathfrak{b}_\alpha(\nu)$ contribute with time oscillations 
$\exp(\pm i(\nu-\omega)t)$. On the time scale $\tau=t/\lambda^2$ and in the weak-coupling limit when $\lambda\to0$, fast oscillations select contributions with $\omega=\nu\geq 0$. Negative transitions frequencies, $\omega\leq 0$, would also be selected if in~\eqref{interaction-h} there were interaction terms of the form $\sigma^{(\alpha)}_+\,\mathfrak{b}^\dag_\alpha(\nu)$ which, together with their Hermitian conjugates, would correspond to the presence of terms of the form $A^\dag_\alpha(\omega)\frak{b}^\dag_\alpha(\nu)$, and Hermitian conjugates, contributing with time oscillations  $\exp(\pm i(\nu+\omega)t)$ .
\end{rem}

\subsection{Spin-chain Hamiltonian: eigenvalues and eigenvectors}
\label{sec:1}

In order to address how the presence of the baths modifies the chain dynamics in the weak-coupling limit and within the global approach, 
we first need diagonalize the chain Hamiltonian in~\eqref{spin-hamiltonian}. By means of the $\ell$-th spin ladder operators in~\eqref{sigmas}
%
%
one rewrites
\begin{equation}
\label{N-Ham1}
H=\Delta\,\sum_{\ell=1}^N\sigma^{(\ell)}_z\,+\,2g\,\sum_{\ell=1}^{N-1}\Big(\sigma^{(\ell)}_+\sigma^{(\ell+1)}_-\,+\,\sigma^{(\ell)}_-\sigma^{(\ell+1)}_+\Big)\ .
\end{equation} 
By means of the 
Jordan-Wigner transformation~\cite{Coleman}, one introduces Fermionic annihilation and creation operators
\begin{equation}
\label{JW1}
a_j:=\prod_{k=1}^{j-1}(-\sigma_z^{(k)})\,\sigma_-^{(j)}\ ,\quad a^\dag_j=\prod_{k=1}^{j-1}(-\sigma_z^{(k)})\,\sigma_+^{(j)}\ ,
\end{equation}
with the convention that $\prod_{k=1}^{j-1}(-\sigma^{(k)}_z)=1$ for $j=1$,
satisfying the anti-commutation relations
\begin{equation}
\label{JW2}
\left\{a_j\,,\,a^\dag_k\right\}=\delta_{jk}\ .
\end{equation}
Let $\vert\uparrow\rangle$ and $\vert\downarrow\rangle$ be the eigenvectors of $\sigma_z$, $\sigma_z\vert\uparrow\rangle=\vert\uparrow\rangle$, 
$\sigma_z\vert\downarrow\rangle=-\vert\downarrow\rangle$. Since $\sigma_-\vert\downarrow\rangle=0$, the vacuum vector such that
$a_j\vert vac\rangle=0$, for all $j=1,2,\ldots,N$, amounts to
\begin{equation}
\label{vac1}
\vert vac\rangle=\vert \downarrow\rangle^{\otimes N}\ .
\end{equation}
Using that
$\displaystyle
a_j^\dag \,a_j=\sigma^{(j)}_+\sigma^{(j)}_-=\frac{1^{(j)}+\sigma^{(j)}_z}{2}$, 
one inverts the transformation~\eqref{JW1}:
\begin{equation}
\label{JW3}
\sigma^{(j)}_-=(\sigma^{(j)}_+)^\dag=\prod_{k=1}^{j-1}\Big(1-2a^\dag_k\,a_k\Big)\,a_j\ ,\quad \sigma^{(j)}_z=2\,a_j^\dag\,a_j\,-\,1\ ,
\end{equation}
%
finally turning the spin Hamiltonian into a Fermionic one, $H=-\,N\,\Delta\,+\,2g\,\widetilde{H}$, where
\begin{equation}
\label{N-Ham3}
\widetilde{H}=\gamma\,\sum_{j=1}^Na^\dag_ja_j\,+\,\sum_{j=1}^{N-1}\Big(a^\dag_j\,a_{j+1}\,+\,a^\dag_{j+1}a_j\Big)\ ,\ \gamma:=\frac{\Delta}{g}\ .
\end{equation} 
As shown in Appendix~\ref{App1}, $H$ can then be diagonalized,
\begin{equation}
\label{N-Ham9}
H=-N\,\Delta+\sum_{\ell=1}^N\left(2\,\Delta+4g\cos\left(\frac{\ell\pi}{N+1}\right)\right)b^\dag_\ell\,b_\ell\ ,
\end{equation}
where the operators 
\begin{equation}
\label{N-Ham8}
b_\ell:=\sum_{j=1}^Nu_{\ell j}\,a_j\ ,\quad b^\dag_\ell:=\sum_{j=1}^Nu_{\ell j}\,a^\dag_j\ ,
\end{equation}
are also Fermionic, $\left\{b_j\,,\,b^\dag_k\right\}=\delta_{jk}$ with the same vacuum as the operators $a_j$: $b_\ell\vert vac\rangle=0$ for all $\ell=1,2,\ldots,N$, while  the coefficients 
\begin{equation}
\label{N-Ham6b}
u_{\ell k}=\sqrt{\frac{2}{N+1}}\sin\left(\frac{\ell k\pi}{N+1}\right)\ .
\end{equation}
form an orthogonal and symmetric matrix $U=[u_{k\ell}]$.

In the following, we shall denote by $\mathbf{n}$ the $N$-tuple $n_1,n_2,\ldots, n_N$, where $n_j=0,1$ is the occupation number of the $j$-th mode relative 
to the operators $b_j$ and $b_j^\dag$. The eigenvectors of the Hamiltonian~\eqref{spin-hamiltonian} have thus the form 
\begin{equation}
\label{N_Ham10}
\vert \mathbf{n}\rangle=(b^\dag_1)^{n_1}(b_2^\dag)^{n_2}\cdots (b_N^\dag)^{n_N}\,\vert vac\rangle\ .
\end{equation}
Indeed, according to~\eqref{N_Ham10},
\begin{eqnarray}
\label{Lindop3a}
b_\ell\vert \mathbf{n}\rangle&=&\delta_{n_\ell,1}\, (-1)^{\sum_{j=1}^{\ell-1}n_j} \sqrt{n_\ell}\,\vert \mathbf{n}_\ell^{-}\rangle\ ,\\
\label{Lindop3b}
b^\dag_\ell\vert \mathbf{n}\rangle&=& \delta_{n_\ell,0}\,(-1)^{\sum_{j=1}^{\ell-1}n_j} \sqrt{1-n_\ell}\,\vert \mathbf{n}^+_\ell\rangle\ ,\\
\label{Lindop3c}
b^\dag_\ell\,b_\ell\vert \mathbf{n}\rangle&=&n_\ell\,\vert \mathbf{n}\rangle\ ,
\end{eqnarray}
where, $\mathbf{n}_\ell^{\pm}$ denote the $N$-tuples $n_1,\ldots,n_\ell\pm1,\ldots,n_N$. Then,
one verifies that $H\,\vert \mathbf{n}\rangle=E_{\mathbf{n}}\,\vert \mathbf{n}\rangle$, where
\begin{equation}
\label{N_Ham11}
E_{\mathbf{n}}=\Delta\Big(2\sum_{\ell=1}^Nn_\ell\,-\,N\Big)+4g\,\sum_{\ell=1}^Nn_\ell\,\cos\left(\frac{\ell\pi}{N+1}\right)\ .
\end{equation}

\begin{rem}
\label{rem1}
Since the matrix $U=[u_{k\ell}]$ with entries $u_{k\ell}$  as in~\eqref{N-Ham6b} is real and symmetric, from~\eqref{N-Ham8} one obtains
\begin{equation}
\label{JW4}
a_j=\sum_{\ell=1}^Nu_{j\ell}\,b_\ell\ ,\quad a^\dag_j=\sum_{\ell=1}^Nu_{j\ell}\,b^\dag_\ell\ .
\end{equation}
Then, by using~\eqref{JW3}, one expresses the Fermionic operators of type $b$ in terms of spin operators:
\begin{eqnarray}
\label{JW5a}
b_\ell&=&\sum_{j=1}^Nu_{\ell j}\,\prod_{k=1}^{j-1}(-\sigma^{(k)}_z)\,\sigma^{(j)}_-\ ,\\
\label{JW5}
b^\dag_\ell&=&\sum_{j=1}^Nu_{\ell j}\,\prod_{k=1}^{j-1}(-\sigma^{(k)}_z)\,\sigma^{(j)}_+\ .
\end{eqnarray}
\end{rem}
A comparison with known results is provided in Appendix~\ref{App_2_3:sec}.

\section{Coupling to external baths}

With the notation of the previous section, the Lindblad operators~\eqref{Lindop1a} now read
\begin{equation}
\label{Lindop1}
A^\dag_\alpha(\omega)=\sum_{E_{\mathbf{m}}-E_{\mathbf{n}}=\omega} \vert\mathbf{m}\rangle\langle \mathbf{m}\vert\,\sigma_+^{(\alpha)}\,
\vert \mathbf{n}\rangle\langle \mathbf{n}\vert\ .
\end{equation}
Their explicit form can be derived by expressing the spin operators $\sigma^{(\alpha)}_+$ 
first in terms of the Fermionic operators $a_j, a^\dag_j$,
\begin{equation}
\label{Lindop2}
\sigma^{(L)}_+=a^\dag_1\ ,\quad
\sigma^{(R)}_+=\prod_{j=1}^{N-1}(1-2\,a^\dag_j\,a_j)\,a_N^\dag\ ,
\end{equation}
and then in terms of the operators $b_\ell, b^\dag_\ell$. Using~\eqref{JW4}, one immediately derives 
\begin{equation}
\label{Lindop4}
\sigma^{(L)}_+=\sum_{\ell=1}^N u_{1\ell}\,b^\dag_\ell\ ,
\end{equation}
while the presence of $\prod_{j=1}^{N-1}(1-2\,a^\dag_j\,a_j)$ in the expression for $\sigma^{(R)}_+$ requires some preliminary 
manipulation.
Firstly, using that $1-2\,a^\dag_j\,a_j=\exp(i\pi\,a^\dag_ja_j)$ and that the  relations~\eqref{JW4} yield
\begin{equation}
\label{exp2}
\sum_{j=1}^{N-1}\,a^\dag_j\,a_j=\sum_{j=1}^Na^\dag_j\,a_j\,-\,a^\dag_N\,a_N=\sum_{\ell=1}^Nb^\dag_\ell\,b_\ell\,-\,a^\dag_N\,a_N\ ,
\end{equation}
one gets
\begin{eqnarray}
\nonumber
{\rm e}^{i\,\pi\,\sum_{j=1}^{N-1}\,a^\dag_j\,a_j}\,a^\dag_N&=&{\rm e}^{i\,\pi\,\sum_{\ell=1}^{N}\,b^\dag_\ell\,b_\ell}\,
{\rm e}^{-i\,\pi\,a^\dag_N\,a_N}\,a^\dag_N\\
\nonumber
&=&{\rm e}^{i\,\pi\,\sum_{\ell=1}^{N}\,b^\dag_\ell\,b_\ell}\,(1\,-\,2a^\dag_N\,a_N)\,a^\dag_N\\
\label{exp3}
&=&-\,{\rm e}^{i\,\pi\,\sum_{\ell=1}^{N}\,b^\dag_\ell\,b_\ell}\,a^\dag_N\ ,
\end{eqnarray}
whence, finally,
\begin{equation}
\label{Lindop5}
\sigma^{(R)}_+=\,-\,\left({\rm e}^{i\,\pi\,\sum_{\ell=1}^{N}\,b^\dag_\ell\,b_\ell}\right)\,\sum_{\ell=1}^Nu_{N\ell}\,b^\dag_\ell\ .
\end{equation}

By means of~\eqref{Lindop4} and~\eqref{Lindop3b}, one then computes the transition amplitudes
\begin{eqnarray}
\label{Lindop6}
\hskip-.7cm
\langle \mathbf{m}\vert\sigma^{(L)}_+\vert \mathbf{n}\rangle=\sum_{\ell=1}^N(-1)^{\sum_{j=1}^{\ell-1}n_j}\,
\sqrt{1-n_\ell}\,u_{1\ell}\,\delta_{\mathbf{m}\mathbf{n}^+_\ell}\ .
\end{eqnarray}
Let $\mathbf{n}_{0_\ell}$ respectively  $\mathbf{n}_{1_\ell}$ denote the $N$-tuples with fixed digits $n_\ell=0$, respectively $n_\ell=1$, at site $\ell$. 
Then, the only contributing transition amplitudes are  
\begin{eqnarray}
\label{Lindop8}
\langle \mathbf{n}_{1_\ell}\vert\sigma^{(L)}_+\vert \mathbf{n}_{0_\ell}\rangle=
(-1)^{\sum_{j=1}^{\ell-1}n_j}\, \sqrt{1-n_\ell}\,u_{1\ell}\ ,
\end{eqnarray}
with $\ell=1,2,\ldots,N$. Also, from~\eqref{N_Ham11} and~\eqref{Lindop1a}, the transition frequencies associated with such amplitudes are
\begin{equation}
\label{Lindop9}
\omega_\ell=E_{\mathbf{n}_{1_\ell}}\,-\,E_{\mathbf{n}_{0_\ell}}
=2\,\Delta\,+\,4\,g\,\cos\left(\frac{\ell\pi}{N+1}\right)\ ,
\end{equation}
while the corresponding Lindblad operators in~\eqref{Lindop1} read
\begin{equation}
\label{Lindop10}
A_L^\dag(\omega_\ell)=u_{1\ell}\,\sum_{\widehat{\mathbf{n}}_{\ell}}(-1)^{\sum_{j=1}^{\ell-1}n_j}\,\vert \mathbf{n}_{1_\ell}\rangle\langle \mathbf{n}_{0_\ell}\vert\ ,
\end{equation}
where the symbol $\sum_{\widehat{\mathbf{n}}_{\ell}}$ means that the summation is performed over all binary $2^{N-1}$-tuples of indices $n_j=0,1$ with $j\neq \ell$.
It thus follows thats
\begin{eqnarray}
\label{SS1a}
A^\dag_L(\omega_\ell)\,A_L(\omega_\ell)&=&u^2_{1\ell}\,\sum_{\widehat{\mathbf{n}}_\ell}\
\vert \mathbf{n}_{1_\ell}\rangle\langle \mathbf{n}_{1_\ell}\vert\ ,\\
\label{SS1c}
A_L(\omega_\ell)\,A^\dag_L(\omega_\ell)&=&u^2_{1\ell}\,\sum_{\widehat{\mathbf{n}}_\ell}\
\vert \mathbf{n}_{0_\ell}\rangle\langle \mathbf{n}_{0_\ell}\vert\ .
\end{eqnarray}
In a similar way, from~\eqref{Lindop5}, one obtains that the only contributing transition amplitudes associated to the right bath are
\begin{equation}
\label{Lindop11a}
\langle \mathbf{n}_{1_\ell}\vert\sigma^{(R)}_+\vert \mathbf{n}_{0_\ell}\rangle=
(-1)^{\sum_{j=1}^{\ell+1}n_j}\, \sqrt{1-n_\ell}\,u_{N\ell}\ ,
\end{equation}
with Lindblad operators
\begin{equation}
\label{Lindop11b}
A_R^\dag(\omega_\ell)=\,u_{N\ell}\,\sum_{\widehat{\mathbf{n}}_\ell}(-1)^{\sum_{j=\ell+1}^Nn_j}\,\vert \mathbf{n}_{1_\ell}\rangle
\langle \mathbf{n}_{0_\ell}\vert\ ,
\end{equation}
whence
\begin{eqnarray}
\label{SS1a}
A^\dag_R(\omega_\ell)\,A_R(\omega_\ell)&=&u^2_{N\ell}\,\sum_{\widehat{\mathbf{n}}_\ell}\
\vert \mathbf{n}_{1_\ell}\rangle\langle \mathbf{n}_{1_\ell}\vert\ ,\\
\label{SS1d}
A_R(\omega_\ell)\,A^\dag_R(\omega_\ell)&=&u^2_{N\ell}\,\sum_{\widehat{\mathbf{n}}_\ell}\
\vert \mathbf{n}_{0_\ell}\rangle\langle \mathbf{n}_{0_\ell}\vert\ .
\end{eqnarray}
%
Notice that,  in the spin representation, all Lindblad operators $A^\dag_\alpha(\omega_\ell)$  involve, through the relations~\eqref{JW3} and~\eqref{JW4}, products of all on-site spin operators. This structure is typical of the global approach to open spin chains and strikingly differs from the local one which yields Lindblad operators involving  
only spin operators pertaining to the first and last spin of the chain.

The operators $A_{\alpha}(\omega_\ell)$ and $A^\dag_\alpha(\omega_\ell)$, $\alpha=L,R$ , have to be inserted into the expressions~\eqref{MS3a} and~\eqref{MS3b} 
when $\omega=\omega_\ell\geq 0$ together with $C^{(\alpha)}_{\omega_\ell}$ and $\widetilde C^{(\alpha)}_{\omega_\ell}$ as in~\eqref{MS4a} and~\eqref{MS4b}.
 Instead, the Lamb-shift Hamiltonian~\eqref{MS5} contributing to~\eqref{MS1} requires the operators $A_\alpha^\dag(\omega_\ell)A_\alpha(\omega_\ell)$ and  
 $A_\alpha(\omega_\ell)A^\dag_\alpha(\omega_\ell)$, $\alpha=L,R$, with both positive and negative $\omega_\ell$.
The Hamiltonian $H+\lambda^2\,H_{LS}$ in~\eqref{MS1} is thus diagonal in the energy eigenbasis $\{\vert\mathbf{n}\rangle\}$.

The $N=2$ and $N=3$ cases are explicitly worked out in Appendix~\ref{app_2_3_open:sec}.

\begin{rem}
\label{rem-imp}
Some observations are in order at this point: the first one is that,  as a consequence of the fact that the transition frequencies contributing to the dissipative generator in~\eqref{MS2} are positive,  not all those corresponding to the non-vanishing
Lindblad operators $A^\dag_\alpha(\omega_\ell)$ in~\eqref{Lindop10} and~\eqref{Lindop11b} need be such.
This means that, the Lindblad operators $A^\dag_\alpha(\omega_\ell)$ with $\omega_\ell<0$ can only contribute to the Lamb-shift Hamiltonian and not to the dissipative part of the generator.
The sign of $\omega_\ell$ depends on the strength of the inter-spin coupling constant $g$; indeed,  
$$
\cos\left(\frac{\pi\ell}{N+1}\right)<0\quad \hbox{for}\quad  N\geq \ell> \frac{N+1}{2}\ .
$$
Correspondingly $\omega_\ell<0$ for
$\displaystyle g>\frac{\Delta}{2\,\left|\cos\left(\frac{\ell\pi}{N+1}\right)\right|}$. \\
The second observation is that, should any of the transition frequencies  $\omega_\ell$ in the list~\eqref{Lindop9} be negative, the opposite one,
$-\omega_\ell=E_{\mathbf{n}_{0_\ell}}-E_{\mathbf{n}_{1_\ell}}$, not being in the list, would not give rise to a dissipative contribution of the form $\mathbb{D}_{-\omega_\ell}[\rho(t)]$, as $A^\dag_\alpha(-\omega_\ell)\equiv 0$.
In the following we shall assume 
\begin{equation}
\label{assump}
g\leq\frac{\Delta}{2\,\cos\left(\frac{\pi}{N+1}\right)}\ ,
\end{equation}
so that $\omega_\ell\geq 0$, $\ell=1,2,\ldots,N$, and leave the study of the presence of negative transition frequencies for future investigations.
\end{rem}

\section{Stationary state}

The master equation~\eqref{MS1} possesses a unique stationary state left invariant by the generated reduced dynamics, namely such that ${\mathbb L}[\rho_\infty]=0$.
This follows from the fact that, as shown in Appendix~\ref{App3}, the only operator commuting with all Lindblad operators $A_{L,R}(\omega_\ell)$ and
$A^\dag_{L,R}(\omega_\ell)$ and with the Hamiltonian must be multiples of the identity~\cite{Spohn2}--~\cite{Fagnola2}.

Because of the diagonal form of $H+\lambda^2H_{LS}$ in the energy eigenbasis, $\Big[H+\lambda^2 H_{LS}\,,\, P_{\mathbf{k}}\Big]=0$, for all
energy eigenprojections $P_{\mathbf{k}}:=\vert \mathbf{k}\rangle\langle \mathbf{k}\vert$.
On the other hand, by inserting $\rho(t)=P_{\mathbf{k}}$ into~\eqref{MS3a} and~\eqref{MS3b}, 
one obtains 
\begin{eqnarray}
\nonumber
&&
\mathbb{D}^{(L)}_{\omega_\ell}[P_{\mathbf{k}}]=u^2_{1\ell}\,\Big(C^{(L)}_{\omega_\ell}\delta_{k_\ell1}-\widetilde{C}^{(L)}_{\omega_\ell}\delta_{k_\ell0}\Big)\times\\
\label{SS3a}
&&\hskip 1cm
\times\,\Big(P_{\mathbf{k}_{0_\ell}}-P_{\mathbf{k}_{1_\ell}}\Big)\\
\nonumber
&&
\mathbb{D}^{(R)}_{\omega_\ell}[P_{\mathbf{k}}]=u^2_{N\ell}\,\Big(C^{(R)}_{\omega_\ell}\delta_{k_\ell1}-\widetilde{C}^{(R)}_{\omega_\ell}\delta_{k_\ell0}\Big)\,\times\\
\label{SS3b}
&&\hskip 1cm
\times\,\Big(P_{\mathbf{k}_{0_\ell}}-P_{\mathbf{k}_{1_\ell}}\Big)\ .
\end{eqnarray}
From~\eqref{MS2}, using the two previous expressions one finds 
\begin{equation}
\label{SS4}
\mathbb{D}\left[P_{\mathbf{k}}\right]=\lambda^2\,\sum_{\ell=1}^N\Big(\delta_{k_\ell1}\,d_\ell\,-\,\delta_{k_\ell0}\,\widetilde{d}_\ell\Big)\,\Big(P_{\mathbf{k}_{0_\ell}}
\,-\,P_{\mathbf{k}_{1_\ell}}\Big)\ ,
\end{equation}
where, using~\eqref{MS4a} and~\eqref{MS4b},
\begin{eqnarray}
\nonumber
\hskip-.5cm
d_\ell&=&C^{(L)}_{\omega_\ell}\,u^2_{1\ell}\,+\,C^{(R)}_{\omega_\ell}\,u^2_{N\ell}\\
\nonumber
\hskip-.5cm
&=&\frac{4\pi\,}{N+1}\sin^2\left(\frac{\pi\ell}{N+1}\right)\,\Big[[h_L(\omega_\ell)]^2\Big(1+n_L(\omega_\ell)\Big)\\
\label{SS5a}
\hskip-1cm
&+&[h_R(\omega_\ell)]^2\Big(1+n_R(\omega_\ell)\Big)\Big]\\
\nonumber
\hskip-1cm
\widetilde{d}_\ell&=&\widetilde{C}^{(L)}_{\omega_\ell}\,u^2_{1\ell}\,+\,\widetilde{C}^{(R)}_{\omega_\ell}\,u^2_{N\ell}\\
\nonumber
\hskip-1cm
&=&\frac{4\pi\,}{N+1}\sin^2\left(\frac{\pi\ell}{N+1}\right)\,
\Big[[h_L(\omega_\ell)]^2\,n_L(\omega_\ell)\,+\\
\hskip-1cm
&+&[h_R(\omega_\ell)]^2\,n_R(\omega_\ell)\Big]\ .
\label{SS5b}
\end{eqnarray}
Consider the diagonal expression 
$X_{diag}=\sum_{\mathbf{n}}x_{\mathbf{n}}\, P_{\mathbf{n}}$; then, the dissipator maps it into
\begin{equation}
\label{SS7}
\mathbb{D}[X_{diag}]=\sum_{\mathbf{n}}\widetilde{x}_{\mathbf{n}}\, P_{\mathbf{n}}\ ,
\end{equation}
with
\begin{equation}
\label{SS7b}
\widetilde{x}_{\mathbf{n}}={\rm Tr}\Big(P_{\mathbf{n}}\,\mathbb{D}[X_{diag}]\Big)=\sum_{\mathbf{m}}\,x_{\mathbf{m}}\,{\rm Tr}\Big(P_{\mathbf{n}}\mathbb{D}\Big[P_{\mathbf{m}}\Big]\Big)\ .
\end{equation}
From~\eqref{SS4} it follows that
\begin{eqnarray}
\nonumber
&&\hskip-.2cm
\mathbb{D}[X_{diag}]=\lambda^2\,\sum_{\mathbf{n}}\, x_{\mathbf{n}}\sum_{\ell=1}^N\Big(\delta_{n_\ell1}\,d_\ell\,-\,\delta_{n_\ell0}
\,\widetilde{d}_\ell\Big)\,\times\\
\label{SS8a}
&&\hskip 1cm
\times\Big(P_{\mathbf{n}_{0_\ell}}\,-\,P_{\mathbf{n}_{1_\ell}}\Big)\,=\\
\label{SS8b}
&&\hskip-.2cm
=\sum_{\ell=1}^N\,\sum_{\hat{\mathbf{n}}_\ell}\Big(d_\ell\,x_{\mathbf{n}_{1_\ell}}\,-\,\widetilde{d}_\ell\,
x_{\mathbf{n}_{0_\ell}}\Big)\,\Big(P_{\mathbf{n}_{0_\ell}}\,-\,P_{\mathbf{n}_{1_\ell}}\Big)\ ,
\end{eqnarray}
whence 
\begin{equation}
\label{SS9}
\widetilde{x}_{\mathbf{n}}=\lambda^2\,\sum_{\ell=1}^N\Big(\delta_{n_\ell0}\,-\,\delta_{n_\ell1}\Big)\,\Big(d_\ell\,
x_{\mathbf{n}_{1_\ell}}\,-\,\widetilde{d}_\ell\,
x_{\mathbf{n}_{0_\ell}}\Big)\ .
\end{equation}
Therefore, $\mathbb{D}[X_{diag}]=0$ is obtained by the factorized expressions
\begin{equation}
\label{SS11}
x_{\mathbf{n}}=\prod_{\ell=1}^N\, x^{(\ell)}_{n_\ell}\ ,\qquad x^{(\ell)}_{0}=d_\ell\ ,\ 
x^{(\ell)}_1=\widetilde{d}_\ell\ .
\end{equation}
All $x^{(\ell)}\geq 0$ and, after normalization, the uniqueness of the stationary state together with the expressions~\eqref{SS5a} and~\eqref{SS5b} yield
\begin{eqnarray}
\label{SS12}
&&
\rho_\infty=\sum_{\mathbf{n}}\Lambda_{\mathbf{n}}\,P_{\mathbf{n}}\ ,\quad
\Lambda_{\mathbf{n}}=\prod_{\ell=1}^N\lambda^{(\ell)}_{n_\ell}\ ,\\ 
\label{SS12c}
&&
\lambda^{(\ell)}_{n_\ell}:=\frac{x^{(\ell)}_{n_\ell}}{x^{(\ell)}_0+x^{(\ell)}_1}=\frac{R^{(\ell)}_{n_\ell}}{R_\ell}\ ,\\
\nonumber
&&
R^{(\ell)}_{n_\ell}:=[h_L(\omega_\ell)]^2\Big(1-n_\ell+n_L(\omega_\ell)\Big)\ +\\
\label{SS12d}
&&\hskip 2cm +\,[h_R(\omega_\ell)]^2\Big(1-n_\ell+n_R(\omega_\ell)\Big)\ ,\\
\nonumber
&&
R_\ell:=[h_L(\omega_\ell)]^2\Big(1+2n_L(\omega_\ell)\Big)\ +\\
\label{SS12e}
&&\hskip 2cm
+\,[h_R(\omega_\ell)]^2\Big(1+2n_R(\omega_\ell)\Big)\ .
\end{eqnarray}
With the simplifying assumption $h_{L,R}(\omega_\ell)=h$, for each $\ell=1,2,\ldots,N$,  one retrieves 
\begin{equation}
\label{SS12b}
\lambda^{(\ell)}_{n_\ell}=\frac{1}{2}\left[1+\frac{(-1)^{n_\ell}}{1+n_L(\omega_\ell)+n_R(\omega_\ell)}\right]\ .
\end{equation}
If we further restrict  to  identical baths, by imposing equal temperatures and thus $\beta_L=\beta_R=\beta$,
one computes
\begin{equation}
\label{Gibbs1}
\lambda^{(\ell)}_{n_\ell}=\frac{{\rm e}^{\beta(1-n_\ell)\omega_\ell}}{{\rm e}^{\beta\omega_\ell}+1}\ ,
\end{equation}
so that
\begin{equation}
\label{Gibbs2}
\rho_\infty=\sum_{\mathbf{n}}\prod_{\ell=1}^N\frac{{\rm e}^{\beta(1-n_\ell)\omega_\ell}}{{\rm e}^{\beta\omega_\ell}+1}\,\vert\mathbf{n}\rangle\langle\mathbf{n}\vert\ .
\end{equation}
On the other hand, using~\eqref{N_Ham11}, 
\begin{equation}
\label{Gibbs3}
\frac{{\rm e}^{-\beta\,E_{\mathbf{n}}}}{\sum_{\mathbf{n}}{\rm e}^{-\beta\,E_{\mathbf{n}}}}=\prod_{\ell=1}^N\frac{{\rm e}^{2\beta(1-n_\ell)\big(\Delta+2g\cos(\frac{\pi\ell}{N+1})\big)}}{
{\rm e}^{2\beta(1-n_\ell)\big(\Delta+2g\cos(\frac{\pi\ell}{N+1})\big)}+1}\ .
\end{equation}
Then,~\eqref{Lindop9} implies that the open chain stationary state $\rho_\infty$ is the Gibbs state at inverse temperature $\beta$ with Hamiltonian $H$ as given 
in~\eqref{spin-hamiltonian}:
\begin{equation}
\label{Gibbs4}
\rho_\infty=\frac{{\rm e}^{-\beta\,H}}{{\rm Tr}\big({\rm e}^{-\beta\,H}\big)}\ .
\end{equation}

\section{Transport properties}

Having determined the explicit, analytic form of the stationary state, we can now study its transport properties
by analyzing the spin and heat flows along he chain, driven by the two external baths.
\vskip .5cm

\subsection{Stationary spin flow: sinks and sources}
\label{subsubsinksources}

The spin flow at site $k=1,2,\ldots,N$ along the spin chain corresponds to the rate of change in time of the average of $\sigma^{(k)}_z$ given by the quantity
\begin{eqnarray}
\nonumber
&&\frac{d}{dt}\text{Tr}\big[ \sigma_z^{(k)}\rho(t)\big]=
\text{Tr}\big[ \sigma_z^{(k)}\mathbb{L}[\rho(t])\big]\\
&&\hskip 3cm
=\text{Tr}\big[\widetilde{\mathbb{L}}[\sigma^{(k)}_z]\,\rho(t)\big]\ .
\label{spin-flux}
\end{eqnarray}
In the first equality $\mathbb{L}$ is the generator at the right hand side of~\eqref{MS1}, while the second equality follows from the cyclicity of the trace, $\text{Tr}(XY)=\text{Tr}(YX)$ and defines the so-called dual generator $\widetilde{\mathbb{L}}$
\begin{equation}
\widetilde{\mathbb{L}}[\sigma^{(k)}_z]= i \big[H+\lambda^2H_{LS},\sigma^{(k)}_z\big]+ \widetilde{\mathbb D}[\sigma^{(k)}_z]\ ,
\label{dual-master-equation}
\end{equation}
with $\widetilde{\mathbb D}[\sigma^{(k)}_z]=\lambda^2 \sum_{\alpha=L,R} \sum_{\omega_\ell\geq 0}^N\ \widetilde{\mathbb D}^{(\alpha)}_{\omega_\ell}[\sigma^{(k)}_z]$, where
\begin{align}
\nonumber
\widetilde{\mathbb D}^{(\alpha)}_{\omega_\ell}[\sigma^{(k)}_z]&=
C^{(\alpha)}_{\omega_\ell} \bigg[ A_\alpha^\dag(\omega_\ell) \sigma^{(k)}_z A_\alpha(\omega_\ell)\\
\label{dualL}
&\hskip 1cm
-\frac{1}{2}\bigg\{A^\dag_\alpha(\omega_\ell) A_\alpha(\omega_\ell), \sigma^{(k)}_z\bigg\}\bigg]\\
\nonumber
&
+\widetilde C^{(\alpha)}_{\omega_\ell} \bigg[ A_\alpha(\omega_\ell) \sigma^{(k)}_z A_\alpha^\dag(\omega_\ell)\\
\label{dualR}
&\hskip 1cm
-\frac{1}{2}\bigg\{A_\alpha(\omega_\ell)A_\alpha^\dag(\omega_\ell), \sigma^{(k)}_z\bigg\}\bigg].
\end{align}

The Hamiltonian contribution to the rate of change in time of the average of $\sigma_z$ 
can be expressed in terms of the dimensionless spin currents:
\begin{align}
J^{(k,k+1)}
=4i\Big(\sigma_-^{(k)}\sigma_+^{(k+1)}-\sigma_+^{(k)}\sigma_-^{(k+1)}\Big)\ ,
\label{spin-current}
\end{align}
as
\begin{equation}
i\Big[H+\lambda^2 H_{LS},\sigma_z^{(k)}\Big]= (g+\kappa)\,\Big(J^{(k-1,k)}-J^{(k,k+1)}\Big)\ ,
\label{current-difference}
\end{equation}
where the Lamb-shift contribution is characterized by a constant
\begin{equation}
\label{current-difference2}
\kappa = \frac{i\lambda^2}{8\sqrt{2}}\sum_{\alpha=L,R}\sum_{\ell=1}^N\,\Big(S_{\omega_\ell}^{(\alpha)}-
\widetilde S_{\omega_\ell}^{(\alpha)}\Big)\ .
\end{equation}
The operator differences in~\eqref{current-difference} thus contribute to the continuity equation~\eqref{spin-flux} as 
current divergence terms with the right dimension of energy.
Since we are interested in the stationary transport properties, we set $\rho(t)=\rho_\infty$ in the right hand side of~\eqref{spin-flux}
and find $\langle J^{(k,k+1)}\rangle_\infty:=\text{Tr}\big(\rho_\infty\, J^{(k,k+1)}\big)=0$.
Indeed, passing from spin to Fermionic operators, by~\eqref{JW3} and~\eqref{JW4}, one finds
\begin{eqnarray}
\nonumber 
J^{(k,k+1)}&=&-4i\big(a_k a^\dag_{k+1}+a_k^\dag a_{k+1}\big)\\
\label{Fermicurr1}
&=&-4i\sum_{j\,\ell=1}^Nu_{k j}u_{k+1 \ell}(b_j b^\dag_\ell+b_j^\dag b_\ell)\ .
\end{eqnarray}
Hence, all their averages with respect to the energy eigenstates vanish,
\begin{equation}
\label{Fermicurr2}
\hskip -.2cm
\langle\mathbf{n}\vert J^{(k,k+1)}\vert\mathbf{n}\rangle=-4i\sum_{j=1}^Nu_{k j}u_{k+1 j}=\langle u_k\vert u_{k+1}\rangle=0\ .
\end{equation}
Indeed,~\eqref{Lindop3a}-\eqref{Lindop3b} yield $\langle\mathbf{n}\vert b_jb^\dag_\ell\vert\mathbf{n}\rangle=\delta_{j\ell}\,(1-n_j)$ and 
$\langle\mathbf{n}\vert b^\dag_jb_\ell\vert\mathbf{n}\rangle=\delta_{j\ell}\,n_j$, while the columns $\vert u_k\rangle$ of the orthogonal and symmetric matrix $U$ (see Remark~\ref{rem1}) are orthogonal.
Thus the stationary left, $\langle J^{(k-1,k)}\rangle_\infty$, and right, $\langle J^{(k,k+1)}\rangle_\infty$, spin  currents through  site $k$ both vanish.

Clearly, being $\rho_\infty$ time-independent, the right hand side of~\eqref{spin-flux} then yields $\text{Tr}\big(\rho_\infty\,\widetilde{\mathbb{D}}[\sigma^{(k)}]\big)=0$.
However, the left and right purely dissipative contributions, 
\begin{equation}
\label{sinksources}
\mathfrak{Q}^{(k)}_{\alpha}:=\lambda^2\sum_{\ell=1}^N\text{Tr}\big(\rho_\infty\,\widetilde{\mathbb{D}}^{(\alpha)}_{\omega_\ell}[\sigma^{(k)}_z]\big)\ ,
\end{equation}
do not separately vanish; indeed, as shown in Appendix~\ref{App5},
\begin{eqnarray}
\nonumber
&&
\mathfrak{Q}^{(k)}_L=2\pi\lambda^2\sum_{\ell=1}^N\,u^2_{k\ell}\,u^2_{1\ell}\,[h_L(\omega_\ell)]^2\,[h_R(\omega_\ell)]^2\,\times\\
\label{sinkssources1}
&&\hskip 2cm
\times\,\frac{n_L(\omega_\ell)-n_R(\omega_\ell)}{R_\ell}\ ,\\
\nonumber
&&
\mathfrak{Q}^{(k)}_R=2\pi\lambda^2\sum_{\ell=1}^N\,u^2_{k\ell}\,u^2_{N\ell}\,[h_L(\omega_\ell)]^2\,[h_R(\omega_\ell)]^2\,\times\\
\label{sinkssources2}
&&\hskip 2cm
\times\,\frac{n_R(\omega_\ell)-n_L(\omega_\ell)}{R_\ell}\ ,
\end{eqnarray}
with $R_\ell$ as in~\eqref{SS12e}. Furthermore, since from~\eqref{N-Ham6b} one finds that $u_{1\ell}=(-)^\ell u_{N\ell}$ for all $\ell=1,2,\ldots,N$, it follows that
$\mathfrak{Q}^{(k)}_L=-\mathfrak{Q}^{(k)}_R$ as it should physically be.

Also, assuming $h_L(\omega)=h_R(\omega)=h$, one gets
\begin{eqnarray}
&&\hskip -.5cm
\label{sinkssources3}
\mathfrak{Q}^{(k)}_L=\pi\lambda^2\sum_{\ell=1}^N\,u^2_{k\ell}\,u^2_{1\ell}\,\frac{n_L(\omega_\ell)-n_R(\omega_\ell)}{1+n_L(\omega_\ell)+n_R(\omega_\ell)}\ ,\\
&&\hskip-.5cm
\label{sinkssources4}
\mathfrak{Q}^{(k)}_R=\pi\lambda^2\sum_{\ell=1}^N\,u^2_{k\ell}\,u^2_{N\ell}\,\frac{n_R(\omega_\ell)-n_L(\omega_\ell)}{1+n_L(\omega_\ell)+n_R(\omega_\ell)}\ .
\end{eqnarray}
One thus sees that, while the continuity equation~\eqref{spin-flux} in the stationary case does not contain any current divergence at site $k$, it does however contain terms of a different origin that are due to the presence of the two baths. These terms vanish only if the temperatures are the same so that $n_L(\omega_\ell)=n_R(\omega_\ell)$ and 
are thus interpretable as spin flow source or sink contributions, depending on whether they are positive or negative.

%
\begin{figure}[h!]
\centering
\includegraphics[width=1\linewidth]{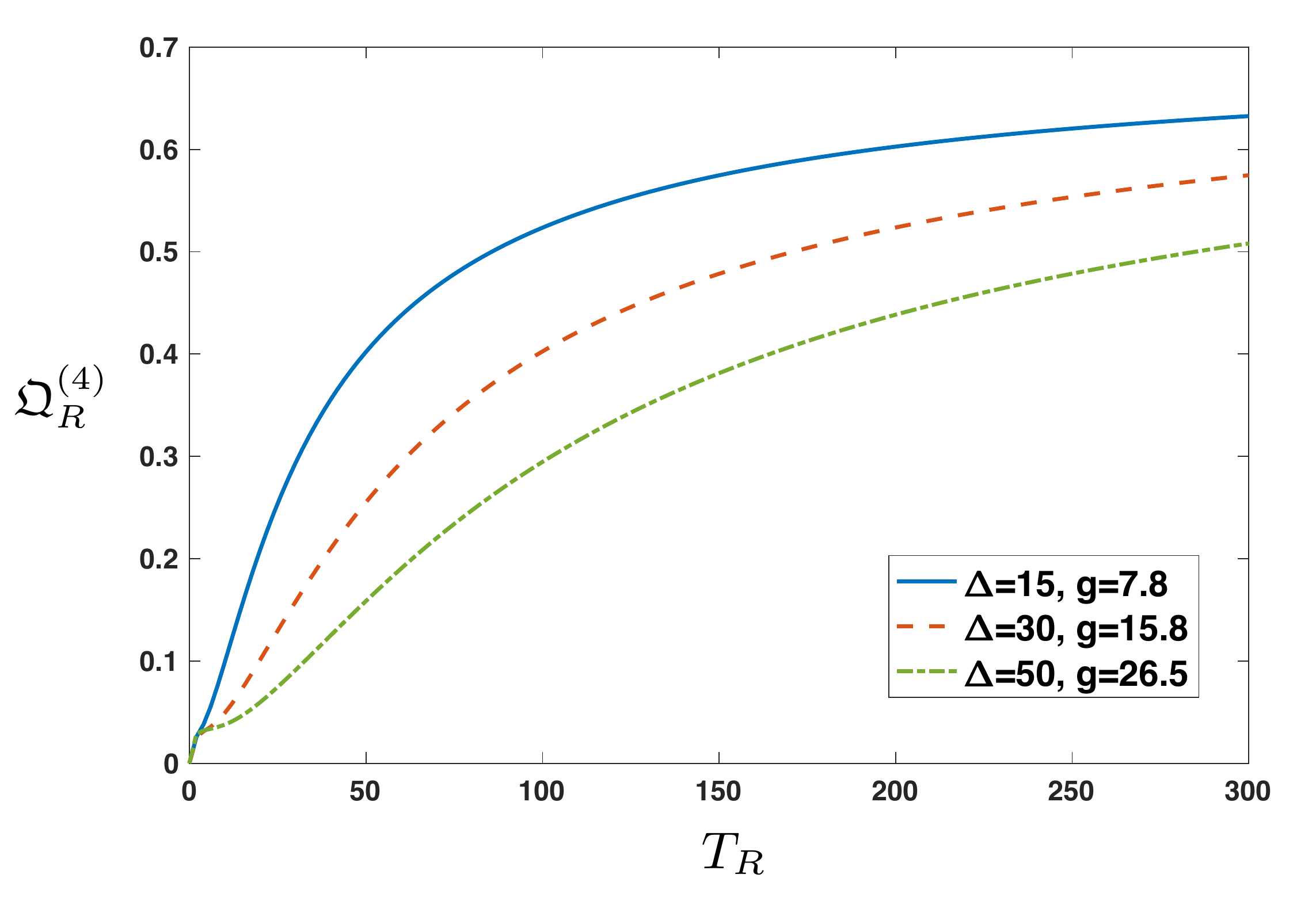}
\caption{Behaviour of the source term $\mathfrak{Q}^{(4)}_R$ as  functions of $T_R$ in a $N=8$ spin chain with left bath temperature $T_L=10$, $\lambda=1$,  $\Delta=15,30,50$ and $g$ close to the saturation values in~\eqref{assump}.}
\label{Fig1:SinkSourceN=5}
\end{figure}

Note that, due to the scaling as $1/N^2$ of the products  $u^2_{k\ell}\,u^2_{N\ell}$ (see~\eqref{N-Ham6b}) and the presence of $N$ of them in~\eqref{sinkssources3} and~\eqref{sinkssources4}, the source and sink terms scale as $1/N$ with increasing  number of spins.
In Figure~\ref{Fig1:SinkSourceN=5} we consider a chain with $N=8$ spins, set $T_L=~0$ so that $n_L(\omega_\ell)=0$ and show the dependence of the source term  
\begin{equation}
\label{SS_TL=0}
\mathfrak{Q}^{(4)}_R=\pi\lambda^2\sum_{\ell=1}^8\,u^2_{4\ell}\,u^2_{8\ell}\,{\rm e}^{-\beta_R\omega_\ell}\ ,
\end{equation}
in the middle of the chain as a function of the right temperature $T_R$ and various values of the transverse magnetic field $\Delta$ and the inter-spin coupling strength $g$.
The values of $g$ associated with $\Delta$ are chosen close to the bound~\eqref{assump},
for reasons that will become clear later when we discuss the bipartite stationary entanglement.


\begin{rem}
\label{rem2a}
The presence of sink and source contributions at sites  $k\neq 1, N$ is strictly related to the global structure of the Lindblad operators in~\eqref{Lindop10} 
and~\eqref{Lindop11b} that involve all spins of the chain.  Should the Lindblad operators depend only on the leftmost and rightmost spin operators as in the local approach to open spin chains (see Remark~\ref{rem-imp}), sink and source terms would disappear as is the case for the two spins in~\cite{Levy}.
Notice that in the global approach developed before sinks and sources are present even in the limit where the inter-spin coupling $g\to0$; indeed, $g$ appears 
in the thermal factors $n_{L,R}(\omega_\ell)$ through the transition frequencies $\omega_\ell$ (see~\eqref{thermnumb}). These terms remain different and non zero 
whenever $\beta_L\neq\beta_R$, even for $g=0$.
\end{rem}

\subsection{Stationary heat flow}
\label{subsubHeatFlow}

Beside the spin flow, the presence of the two baths at the far ends of the chain also establishes heat flows in and out of the chain.
According to standard quantum thermodynamics arguments~\cite{Alicki1,Spohn1}, the heat flow through an open quantum system due to its weak coupling to a
thermal bath, is measured by 
\begin{equation}
\label{HF1}
\mathfrak{H}(t):=\text{Tr}\left(\frac{{\rm d}\rho(t)}{{\rm d}t}\,H\right)=\text{Tr}\left(\mathbb{L}[\rho(t)]\,H\right)\ ,
\end{equation} 
where $\rho\mapsto\rho(t)$ is the dissipative evolution due to the bath, generated by $\mathbb{L}$, while $H$ is the open system time-independent Hamiltonian.  
Because of the structure of  the GKSL equation as in~\eqref{MS1}, only the dissipative term of the generator contributes to the heat flow; therefore, in the spin chain stationary state, the heat flow due to the left, respectively right bath is given by 
\begin{equation}
\label{HF2}
\mathfrak{H}^{st}_{\alpha}=\sum_{\ell=1}^N\text{Tr}\left(\mathbb{D}_{\omega_\ell}^{(\alpha)}[\rho_\infty]\,H\right)\ , \quad\alpha=L,R\ .
\end{equation} 
Certainly, $\mathbb{D}[\rho_\infty]=0$ implies $\mathfrak{H}^{st}_L+\mathfrak{H}^{st}_R=0$; however, as for the spin flow, the single bath contributions to the heat flow need not separately vanish and their sign, if positive, corresponds to heat flowing into the chain from the bath, or to heat flowing out of the chain and into the bath.

\begin{figure}[h!]
 \centering
\includegraphics[width=1\linewidth]{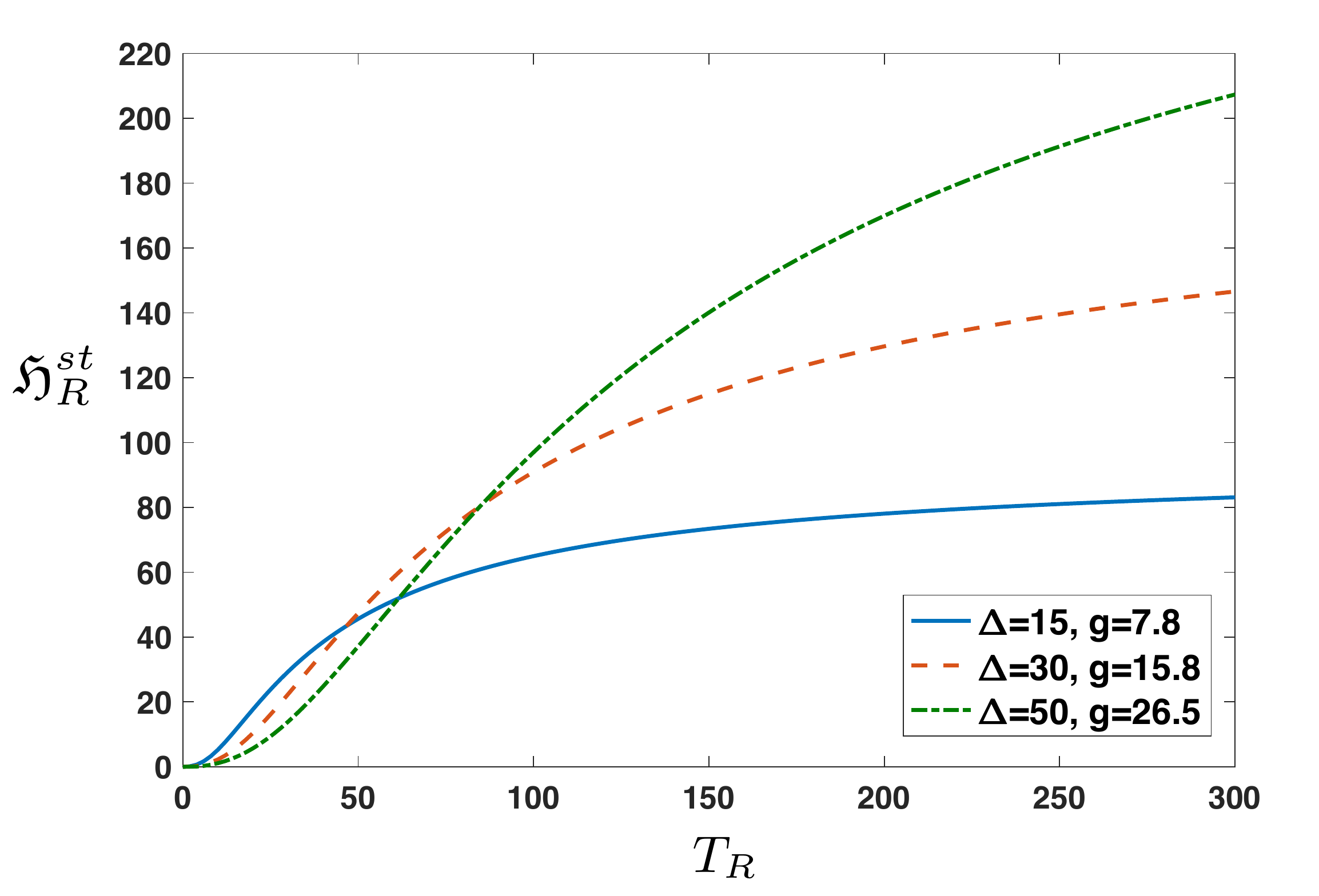}
\caption{Behaviour of the heat flow $\mathfrak{H}^{st}_R$ as a function of the right bath temperature $T_R$ in a $N=8$ spin chain with left bath temperature $T_L=0$, $\lambda=1$,
$\Delta=15,30,50$ and $g$ close to the corresponding saturation values.}
\label{Fig2:HeatFlowN=5}
\end{figure}

Using~\eqref{Lindop9},~\eqref{SS3a},~\eqref{SS12}-\eqref{SS12e},~\eqref{MS4a} and~\eqref{MS4b} one computes
\begin{eqnarray}
\nonumber
&&
\mathfrak{H}^{st}_L=\sum_{\ell=1}^N\text{Tr}\left(\mathbb{D}^{(L)}_{\omega_\ell}[\rho_\infty]\,H\right)=\lambda^2\sum_{\ell=1}^N\sum_{\mathbf{n},\mathbf{k}}\, 
E_{\mathbf{n}}\,\Lambda_{\mathbf{k}}\,u^2_{1\ell}\,\times\\
\nonumber
&&\hskip 1cm
\times\,\Big(C^{(L)}_{\omega_\ell}\delta_{1k_\ell}-\widetilde{C}^{(L)}_{\omega_\ell}\delta_{0k_\ell}\Big)\,
\langle\mathbf{n}\vert\Big(P_{\mathbf{k}_{0_\ell}}-P_{\mathbf{k}_{1_\ell}}\Big)\vert\mathbf{n}\rangle\\
\nonumber
&&
\hskip .5cm
=\lambda^2\sum_{\ell=1}^N\omega_\ell\,u^2_{1\ell}\, \Big(C^{(L)}_{\omega_\ell}\,\lambda^{(\ell)}_1\,-\,\widetilde{C}^{(L)}_{\omega_\ell}\,\lambda^{(\ell)}_0\Big)\\
\nonumber
&&
\hskip .5cm
=\lambda^2\sum_{\ell=1}^N\omega_\ell\,u^2_{1\ell}\, [h_L(\omega_\ell)]^2[h_R(\omega_\ell)]^2\,\times\\
\label{HF3}
&&
\hskip 1cm
\times\,\frac{n_L(\omega_\ell)-n_R(\omega_\ell)}{R_\ell}\ .
\end{eqnarray}
Notice that the heat flow is positive, namely it flows from the left bath into the chain if $n_L(\omega)> n_R(\omega)$, that is (see~\eqref{thermnumb})  if the left bath is at higher temperature than the right one.
Furthermore, the simplifying assumption $h_L(\omega)=h_R(\omega)=h$ yields
\begin{equation}
\label{HF4}
\mathfrak{H}^{st}_R=\pi\,\lambda^2\,\sum_{\ell=1}^N\omega_\ell\,u^2_{1\ell}\,\frac{n_R(\omega_\ell)-n_L(\omega_\ell)}{1+n_L(\omega_\ell)+n_R(\omega_\ell)}\ .
\end{equation}
%
Furthermore,  the transition frequencies $\omega_\ell$ in~\eqref{Lindop9} are of order $1$ with respect to increasing $N$, whence each of the $N$ contributions $\omega_\ell\, u^2_{1\ell}$ to the heat flow scales as $1/N$ due to~\eqref{N-Ham6b}. Thus, unlike the sink and source terms in~\eqref{sinkssources3} and~\eqref{sinkssources4} that scale as $1/N$, the heat flow does not vanish with increasing $N$.
Setting $N=8$ and $T_L=0$ as for the source terms in~\eqref{SS_TL=0}, and choosing the same set of parameters $\Delta$ and $g$ as in Figure~\ref{Fig1:SinkSourceN=5}, the behaviour of the heat flow $\mathfrak{H}^{st}_R$ as a function of $T_R$ is reported in Figure~\ref{Fig2:HeatFlowN=5}.
%

\section{Entanglement properties}

Besides transport phenomena, open spin chains represent attractive models of many-body systems due to their entanglement properties. Indeed, although the external, transverse magnetic field tends to align all spins in a separable state, the inter-spin interaction instead is able to generate quantum correlations among all spins.
The presence of the external baths at the chain end points constitutes interesting additional driving factors
influencing the behaviour of the spin entanglement.

In what follows we shall focus upon the entanglement between any two spins, $r$ and $s> r$, in the stationary state via the concurrence of the reduced 
bipartite density matrix $\rho_{r,s}$ obtained from~\eqref{SS12} by tracing over the spins at sites different from  $r$ and $s$.
In order to achieve this goal, one needs 
to re-express the stationary state in~\eqref{SS12}  in terms of spin operators, rather than Fermionic ones.
\vskip .5cm

\subsection{Stationary state: spin representation}
\label{ststsprep:sec}


In this respect, instead of the standard lexicograhic  ordering, it proves convenient to regroup the binary strings $\mathbf{n}$ in terms of the number of ones they contain. We then introduce the enumeration of the $2^N$ binary $N$-tuples $\mathbf{n}$ known as combinatorial numbering of degree $p$~\cite{combinadic}, that we shall refer to as combinadic ordering for sake of 
shortness. 
For fixed $p=0,1,2,\ldots,N$, one bijiectively associates to each of the
$N \choose p$ $N$-tuples with $n_{i_\ell}=1$ at sites  $i_1<\cdots<i_p$ the integers 
\begin{equation}
\label{combenum}
1\leq \mathcal{N}_p=\,1\,+\,\sum_{\ell=1}^p {i_\ell-1\choose \ell}\leq {N\choose p}\ ,
\end{equation}
where the binomial coefficients ${i_\ell-1\choose \ell}$ are set to vanish if $i_\ell-1<\ell$. 
According to such a numbering, we identify $\mathbf{n}$ with a unique $\mathcal{N}_p$ for some $p=0,1,\ldots,N$; then,
the stationary state $\rho_\infty$ may be written as 
\begin{equation}
\label{SS12a}
\rho_\infty=\sum_{p=0}^N\sum_{\mathcal{N}_p=1}^{{N\choose p}}\mathcal{L}^{(p)}_{\mathcal{N}_p}\,\vert\mathcal{N}_p\rangle\langle\mathcal{N}_p\vert\ ,
\end{equation}
where $\displaystyle\mathcal{L}^{(p)}_{\mathcal{N}_p}$ denotes the eigenvalue $\Lambda_{\mathbf{n}}$ in~\eqref{SS12} corresponding to the binary $N$-tuple  $\mathbf{n}$ with $p$ 1's, indexed by the combinadic 
integer $\mathcal{N}_p$.
Applications of the above formalism to the $N=2$ and $N=3$ cases can be found in Appendix~\ref{app_2_3_stst:sec}.

Notice that, for any fixed $p=0,1,\cdots,N$, the integers $\mathcal{N}_p$ in~\eqref{combenum} correspond 
to the  Fermionic excitations of the modes $i_1<\cdots< i_p$ of type $b$. Indeed,  $\mathcal{N}_p$ identifies  the binary $N$-tuple $\mathbf{n}$, where $n_{i_1}=\cdots=n_{i_p}=1$ while the remaining $n_j$ vanish. 
We can thus consistently label:
\begin{equation}
\label{XSC0a}
\vert \mathcal{N}_p\rangle=b_{i_1}^\dag\cdots b^\dag_{i_p}\vert vac\rangle\ .
\end{equation}
Then, using~\eqref{N-Ham8} one writes
\begin{equation}
\label{XSCV1a}
\vert \mathcal{N}_p\rangle=\sum_{j_1,j_2,\ldots,j_p}u_{i_1j_1}u_{i_2j_2}\cdots u_{i_pj_p}\, a^\dag_{j_1}a^\dag_{j_2}\cdots a^\dag_{j_p}\vert vac\rangle\ .
\end{equation}
Notice that, unlike the indices $i_1,\ldots,i_p$,  the indices $j_1,j_2,\ldots, j_p$ are in general not ordered; their reordering such that $j_1<j_2<\cdots<j_p$ yields
\begin{equation}
\label{XSCV1b}
\vert \mathcal{N}_p\rangle=\sum_{j_1<j_2<\ldots<j_p}D^{i_1<\cdots< i_p}_{j_1<\cdots<j_p}\, a^\dag_{j_1}a^\dag_{j_2}\cdots a^\dag_{j_p}\vert vac\rangle\ ,
\end{equation}
where 
\begin{equation}
\label{XSC2}
D^{i_1<\cdots< i_p}_{j_1<\cdots<j_p}=\hbox{det}\Big(U^{i_1<\cdots<i_p}_{j_1<\cdots<j_p}\Big)\ ,
\end{equation}
is the determinant of the $p\times p$ 
sub-matrix $U^{i_1<\cdots<i_p}_{j_1<\cdots<j_p}$  of the orthogonal and symmetric matrix $U$ in Remark~\ref{rem1}  with $p$ rows indexed by $i_1<\cdots<i_p$ and $p$ columns by $j_1<\cdots<j_p$. Its entries are thus given by 
\begin{equation}
\label{XSC2b}
\left(U^{i_1<\cdots<i_p}_{j_1<\cdots<j_p}\right)_{ab}=u_{i_aj_b}=\sqrt{\frac{2}{N+1}}\,\sin\Big(\frac{\pi\,i_aj_b}{N+1}\Big)\ .
\end{equation}

\begin{rem}
\label{rem5}
For $p=0$, all $n_j=0$ whence there are no $i_\ell$ and $j_\ell$ to choose and one sets $U^{i_1<\cdots<i_p}_{j_1<\cdots<j_p}=1$.
Instead, if only $n_{i_\ell}=1$ then the matrices 
$U^{i_1<\cdots<i_p}_{j_1<\cdots<j_p}$ reduce to the scalars $U^{i_\ell}_{j_\ell}=u_{i_\ell j_\ell}$. Finally, if all $n_j=1$, then there is only one contributing matrix,
$U^{12\cdots N}_{12\cdots N}=U$, and ${\rm Det}(U)=-1$. Unlike the matrix $U$ the  sub-matrices $U^{i_1<\cdots<i_p}_{j_1<\cdots<j_p}$ are not symmetric; however,
\begin{equation}
\label{XSC2a}
D^{i_1<\cdots<i_p}_{j_1<\cdots<j_p}
=D^{j_1<\cdots<j_p}_{i_1<\cdots<i_p}\ .
\end{equation}
It is convenient to introduce the ${N\choose p}\times{N\choose p}$ matrices $\mathcal{D}^{(p)}$, where $\mathcal{D}^{(0)}=1$ and $\mathcal{D}^{(N)}=-1$ are
scalars, otherwise $\mathcal{D}^{(p)}$ has entries $\mathcal{D}^{(p)}_{\mathcal{N}'_p\mathcal{N}''_p}$ 
corresponding to the determinants 
$D^{i'_1<\cdots< i'_p}_{i''_1<\cdots<i''_p}$, where $\mathcal{N}'_p$ identifies an $N$-tuple with $1$'s at sites $i'_1<\cdots<i'_p$, $\mathcal{N}''_p$ identifies an $N$-tuple 
with $1$'s at sites $i''_1<\cdots<i''_p$. Because of~\eqref{XSC2a}, the matrices $\mathcal{D}^{(p)}$ are symmetric for all $p=0,1,\ldots,N$.
\end{rem}

In the spin representation $\vert vac\rangle=\vert\downarrow\rangle^{\otimes N}$; therefore, as shown in Appendix~\ref{App6}, 
setting $\vert\uparrow\rangle=\vert1\rangle_S$ and 
$\vert\downarrow\rangle=\vert 0\rangle_S$ so that  $\sigma_+\vert 0\rangle_S=\vert 1\rangle_S$,
one can express the Fermionic states $\vert\mathcal{N}_p\rangle$ with $p$ excitations at sites $i_1<\cdots<i_p$ as linear combinations of the spin states 
$\vert \mathcal{N}'_p\rangle_S$ with $p$ spins flipped up at the sites 
$i'_1<\cdots<i'_p$ identified by the combinadic index $\mathcal{N}'_p$. It follows that, with respect to  the standard spin basis, the stationary state 
$\rho_\infty$ in~\eqref{SS12} can be recast as
\begin{eqnarray}
\label{XSC9c0}
\rho_\infty&=&\sum_{p=0}^N\,\sum_{\mathcal{N}'_p,\mathcal{N}''_p}\,\mathcal{S}^{(p)}_{\mathcal{N}'_p\mathcal{N'}'_p}\,\vert \mathcal{N}'_p\rangle_S{}_S\langle\mathcal{N}''_p\vert\ ,\\
\label{XSC9c}
\mathcal{S}^{(p)}_{\mathcal{N}'_p\mathcal{N}''_p}&:=&\sum_{\mathcal{N}_p}\,\mathcal{L}_{\mathcal{N}_p}\,
\mathcal{D}^{(p)}_{\mathcal{N}_p\mathcal{N}'_p}\,\mathcal{D}^{(p)}_{\mathcal{N}_p\mathcal{N}''_p}\ ,
\end{eqnarray}
where $\displaystyle\mathcal{L}_{\mathcal{N}_p}$ are the eigenvalues of $\rho_\infty$ as in~\eqref{SS12}.

From~\eqref{XSC2a} and~\eqref{XSC9c} it follows that, in the standard spin basis, the stationary state is represented by the block-diagonal matrix 
\begin{equation}
\label{blockS}
\mathcal{S}=\bigoplus_{p=0}^N\mathcal{S}^{(p)}\ ,\ \mathcal{S}^{(p)}={\mathcal D}^{(p)}\,{\mathcal L}^{(p)}\,{\mathcal D}^{(p)}\ ,\ 
\mathcal{D}=\bigoplus_{p=0}^N\mathcal{D}^{(p)}\ ,
\end{equation}
where $\mathcal{L}^{(p)}$ is the diagonal matrix whose entries are the eigenvalues in~\eqref{SS12}  labelled by the combinadic integers  
$\mathcal{N}_p$ while the matrices $\mathcal{D}^{(p)}$  are as in the previous remark.

Finally, again in Appendix~\ref{App6} it is shown that the stationary state $\rho_\infty$ has the following structure in terms of spin operators
\begin{equation}
\label{XSC9a}
\rho_\infty=\sum_{p=0}^N\,\sum_{\mathcal{N}'_p,\mathcal{N}''_p}\,\mathcal{S}^{(p)}_{\mathcal{N}'_p\mathcal{N}''_p}\,\prod_{\ell=1}^N\left(X_{n'_\ell}^{(\ell)}\left(X_{n''_\ell}^{(\ell)}\right)^\dag\right)\ ,
\end{equation}
where $\displaystyle X_0^{(\ell)}=\frac{1-\sigma^{(\ell)}_z}{2}$ and $X^{(\ell)}_1=\sigma_+^{(\ell)}$, while $n'_\ell$ and $n''_\ell$ are the digits of the binary $N$-tuples with
combinadic indices $\mathcal{N}'_p$ and $\mathcal{N}''_p$. 

The above expression of the stationary state $\rho_\infty$ can thus be algorithmically computed for any $N$; the cases $N=2,3$ provide concrete and informative analytical instances of the above structure and are dealt with in Appendix~\ref{App7}. 

\subsection{Two-spin entanglement}

The spin-operator expression of the stationary state $\rho_\infty$ is useful to investigate the entanglement content of any pair of spins along the chain and its dependence on 
their positions $N\geq s>r\geq 1$.
We shall quantify the two-spin entanglement by means of the concurrence~\cite{Wootters} of the two-spin reduced density matrix which is obtained by tracing over
the spins at sites different from $r$ and $s$, operation that will be denoted  as ${\rm Tr}_{(r,s)}$.
Considering the expression~\eqref{XSC9a} one readily computes
\begin{eqnarray}
\nonumber
&&\hskip -.5cm
{\rm Tr}_{(r,s)}\left(\prod_{\ell=1}^N\left(X_{n'_\ell}^{(\ell)}\left(X_{n''_\ell}^{(\ell)}\right)^\dag\right)\right)=\\
\nonumber
&&\hskip .5cm
=\left(\prod_{\ell\neq r,\ell\neq s}{\rm Tr}\left(X_{n'_\ell}^{(\ell)}\,\left(X_{n''_\ell}^{(\ell)}\right)^\dag\right)\right)\times\\
\nonumber
&&\hskip -.5cm
\times\,\left(X_{n'_r}^{(r)}\,\left(X^{(r)}_{n''_r}\right)^\dag\right)\, \left(X_{n'_s}^{(s)}\left(X^{(s)}_{n''_s}\right)^\dag\right)\\
\label{XSC10}
&&\hskip -.5cm
=\left(\prod_{\ell\neq r,\ell\neq s}\delta_{n'_\ell n''_\ell}\right)\ \left(X_{n'_r}\,X^\dag_{n''_r}\right)\,\otimes \left(X_{n'_s}\,X^\dag_{n''_s}\right)\ ,
\end{eqnarray}
where, in the final two-spin expression, the reference to the spin sites has safely been neglected.
We thus see that the partial trace reduces the double sum over all possible binary strings $\mathbf{n}'$ and $\mathbf{n}''$ 
in~\eqref{XSC9a} to a double sum over binary strings that have equal digits but, possibly, for the sites $r$ and $s$. We shall then denote by $\mathcal{N}^{(rs)}_p(n'_r,n'_s)$ 
and $\mathcal{N}^{(rs)}_p(n''_r,n''_s)$ the combinadic inidices~\eqref{combenum} of the binary strings with $p$ ones that have the same entries $n_j$ everywhere but, 
possibly, for the sites $r$ and $s$. 

With this notation, the two-spin density matrix formally reads
\begin{eqnarray}
\nonumber
\rho_{(r,s)}&:=&{\rm Tr}_{(r,s)}\big(\rho_\infty\big)\\
\nonumber
&=&\sum_{\substack{n'_r,n'_s\\ n''_r,n''_s}}\,\sum_{p=0}^N\sum_{\substack{\mathcal{N}^{(rs)}_p(n'_r,n'_s)\\
\mathcal{N}^{(rs)}_p(n''_r,n''_s)}}\,\mathcal{S}^{(p)}_{\mathcal{N}^{(rs)}_p(n'_r,n'_s)\mathcal{N}^{(rs)}_p(n''_r,n''_s)}\times\\
\label{XSC11}
&\times& \left(X_{n'_r}\,X^\dag_{n''_r}\right)\,\otimes \left(X_{n'_s}\,X^\dag_{n''_s}\right)\ .
\end{eqnarray}
Since the $N$-tuples indexed by $\mathcal{N}^{(rs)}_p(n''_r,n''_s)$ have the same entries but, possibly, for the sites $r$ and $s$, it follows that the allowed values for $n'_r,n'_s$ and $n''_r,n''_s$ must satisfy 
$n'_r\,+\,n'_s=n''_r\,+\,n''_s$. These latter ones and the corresponding two-spin operators are as follows:
\begin{align}
\label{XSC12a1}
&
\left\{\begin{matrix}
n'_r=0\ ,&n''_r=0\cr
n'_s=0\ ,&n''_s=0\end{matrix}\right. :\quad
\frac{1-\sigma_z}{2}\otimes\frac{1-\sigma_z}{2}\ ,\\
\label{XSC12a2}
&
\left\{\begin{matrix}
n'_r=0\ ,&n''_r=0\cr
n'_s=1\ ,&n''_s=1\end{matrix}\right. :\quad
\frac{1-\sigma_z}{2}\otimes\frac{1+\sigma_z}{2}\ ,
\end{align}
\begin{align}
\label{XSC12b1}
&
\left\{\begin{matrix}
n'_r=0\ ,&n''_r=1\cr
n'_s=1\ ,&n''_s=0\end{matrix}\right. :\quad
\sigma_-\otimes\sigma_+\ ,\\
\label{XSC12b2}
&
\left\{\begin{matrix}
n'_r=1\ ,&n''_r=1\cr
n'_s=0\ ,&n''_s=0\end{matrix}\right. :\quad \frac{1+\sigma_z}{2}\otimes\frac{1-\sigma_z}{2}\ ,
\end{align}
and
\begin{align}
\label{XSC12c1}
&
\left\{\begin{matrix}
n'_r=1\ ,&n''_r=0\cr
n'_s=0\ ,&n''_s=1\end{matrix}\right. :\quad
\sigma_+\otimes\sigma_-\ ,\\
\label{XSC12c2}&
\left\{\begin{matrix}
n'_r=1\ ,&n''_r=1\cr
n'_s=1\ , &n''_s=1\end{matrix}\right. :\quad
\frac{1+\sigma_z}{2}\otimes\frac{1+\sigma_z}{2}\ .
\end{align}
Therefore, for all sites $1\leq r<s\leq N$, the reduced two-spin density matrix is a $X$-state 
for the case of a two-spin chain (see~\eqref{2spinspinrep3} in Appendix~\ref{App7}):
\begin{eqnarray}
\label{XSC13a}
\hskip-.8cm
\rho_{(r,s)}&=&\begin{pmatrix}a&0&0&0\cr
0&b&c&0\cr
0&c&d&0\cr
0&0&0&e
\end{pmatrix}\\
\\
\hskip-.8cm
&=&
a\,\frac{1+\sigma_z}{2}\otimes\frac{1+\sigma_z}{2}\,+\,b\,\frac{1+\sigma_z}{2}\otimes\frac{1-\sigma_z}{2}\\
\label{XSC13c}
\hskip-.8cm
&+&
c\,\sigma_+\otimes\sigma_-
\,+\,c\,\sigma_-\otimes\sigma_+\\
\label{XSC13b}
\hskip-.8cm
&+&d\,\frac{1-\sigma_z}{2}\otimes\frac{1+\sigma_z}{2}\,+\,e\,\frac{1-\sigma_z}{2}\otimes\frac{1-\sigma_z}{2}\ ,
\end{eqnarray}
with off-diagonal entry 
\begin{equation}
\label{XSC14b}
c=\sum_{p=0}^N\sum_{\substack{\mathcal{N}^{(rs)}_p(1,0)\\ \mathcal{N}^{(rs)}_p(0,1)}}\,\mathcal{S}^{(p)}_{\mathcal{N}^{(rs)}_p(1,0)\mathcal{N}^{(rs)}_p(0,1)}\ .
\end{equation}
Here the combinadic indices of the entries of $\mathcal{S}^{(p)}$ contributing to the only off-diagonal term $c$ involve different sites $r$ and $s$.
The combinadic indices  are instead the same for the entries of $\mathcal{S}^{(p)}$ contributing to the diagonal entries:
\begin{eqnarray}
\label{XSC14a}
a&=&\sum_{p=0}^N\sum_{\mathcal{N}^{(rs)}_p(1,1)}\,\mathcal{S}^{(p)}_{\mathcal{N}^{(rs)}_p(1,1)\mathcal{N}^{(rs)}_p(1,1)}\ ,\\
\label{XSC14a1}
b&=&\sum_{p=0}^N\sum_{\mathcal{N}^{(rs)}_p(1,0)}\,\mathcal{S}^{(p)}_{\mathcal{N}^{(rs)}_p(1,0)\mathcal{N}^{(rs)}_p(1,0)\ ,}\\
\label{XSC14c}
d&=&\sum_{p=0}^N\sum_{\mathcal{N}^{(rs)}_p(0,1)}\,\mathcal{S}^{(p)}_{\mathcal{N}^{(rs)}_p(0,1)\mathcal{N}^{(rs)}_p(0,1)}\ ,\\
\label{XSC14c1}
e&=&\sum_{p=0}^N\sum_{\mathcal{N}^{(rs)}_p(0,0)}\,\mathcal{S}^{(p)}_{\mathcal{N}^{(rs)}_p(0,0)\mathcal{N}^{(rs)}_p(0,0)}\ .
\end{eqnarray}
For such states the concurrence takes the following analytic expression
\begin{equation}
\label{XSC15}
C(r,s)=2\,\max\left\{0\,,\,\Big(|c|-\sqrt{a\,e}\Big)\right\}\ .
\end{equation}
whence the stationary bipartite entanglement corresponding to a non-vanishing positive $C(r,s)$, can be evaluated as a function of the sites $r$ and $s$ and their distance $s-r$.
The concurrences $C(1,2)$, $C(2,3)$ and $C(1,3)$ for a three spin chain are studied in Appendix~\ref{App9}. 
\vskip .5cm


\subsection{Two-spin concurrence}
\label{Concurrence}

In this section we study the stationary two-spin entanglement in a $N$-spin chain. In doing so, we use Appendix~\ref{App8} which shows how the coefficients $a,b,c,d$ and $e$ appearing in the concurrence $C(r,s)$ in~\eqref{XSC15} can be algorithmically reconstructed.
The quantity $C(r,s)$ depends on  the parameters $\Delta$ and $g$ of the chain Hamiltonian, on the temperatures $T_{L,R}$, on the number  of spins, $N$, and on the spin sites 
$0\leq r\leq s\leq N$.

Firstly, although the algorithm developed in Appendix~\ref{App8} 
works for all $N$, its algorithmic implementation rapidly becomes time-consuming so that, in the following figures, we shall focus upon a chain consisting of $N=8$ spins.
In full generality, we observe that, similarly to the sink and source terms in~\eqref{sinkssources1}, and~\eqref{sinkssources2}, the bipartite entanglement between any pair of sites scales as $1/N$; this follows from the fact that, for large $N$, such is the leading order of the matrix elements $\mathcal{S}^{(p)}_{\mathcal{N}'_p\mathcal{N}''_p}$ in~\eqref{XSC9c}. In turn, such a behaviour is due to the fact that the transition frequencies $\omega_\ell$ in~\eqref{Lindop9}, and thus the eigenvalues~\eqref{SS12}, are of order $1$ with respect to $N$, while the quantities $\mathcal{D}^{(p)}_{\mathcal{N}'_p\mathcal{N}''_p}$ introduced in Remark~\ref{rem5} are of order $1/(\sqrt{N})^p$ and,  in each of the expressions~\eqref{XSC14b}--\eqref{XSC14c1}, there appear sums from $p=1$ to $p=N$ of products of pairs of such terms.

Secondly, as much as in the case of  source and sink terms and of heat flows, we set $T_L=0$ and then inspect the dependence on the right temperature $T_R$ only.
What one expects by letting $T_L>0$ is that when $T_R=T_L>0$ one reaches the Gibbs state in~\eqref{Gibbs4}. This thermal equilibrium  state can not provide transport effects, for $n_L(\omega_\ell)=n_R(\omega_\ell)$, but may however support bipartite entanglement at finite non-vanishing temperatures. On the other hand,  for $T_L=T_R=0$ the state becomes the vacuum state $\vert vac\rangle$ in~\eqref{vac1} which is clearly separable.
close to the maximum value~\eqref{assump} that ensure the positivity of all transition frequencies $\omega_\ell$ in~\eqref{Lindop9}))  and plot various concurrences versus $T_R$. 

\begin{figure}[h!]
\centering
\includegraphics[width=\linewidth]{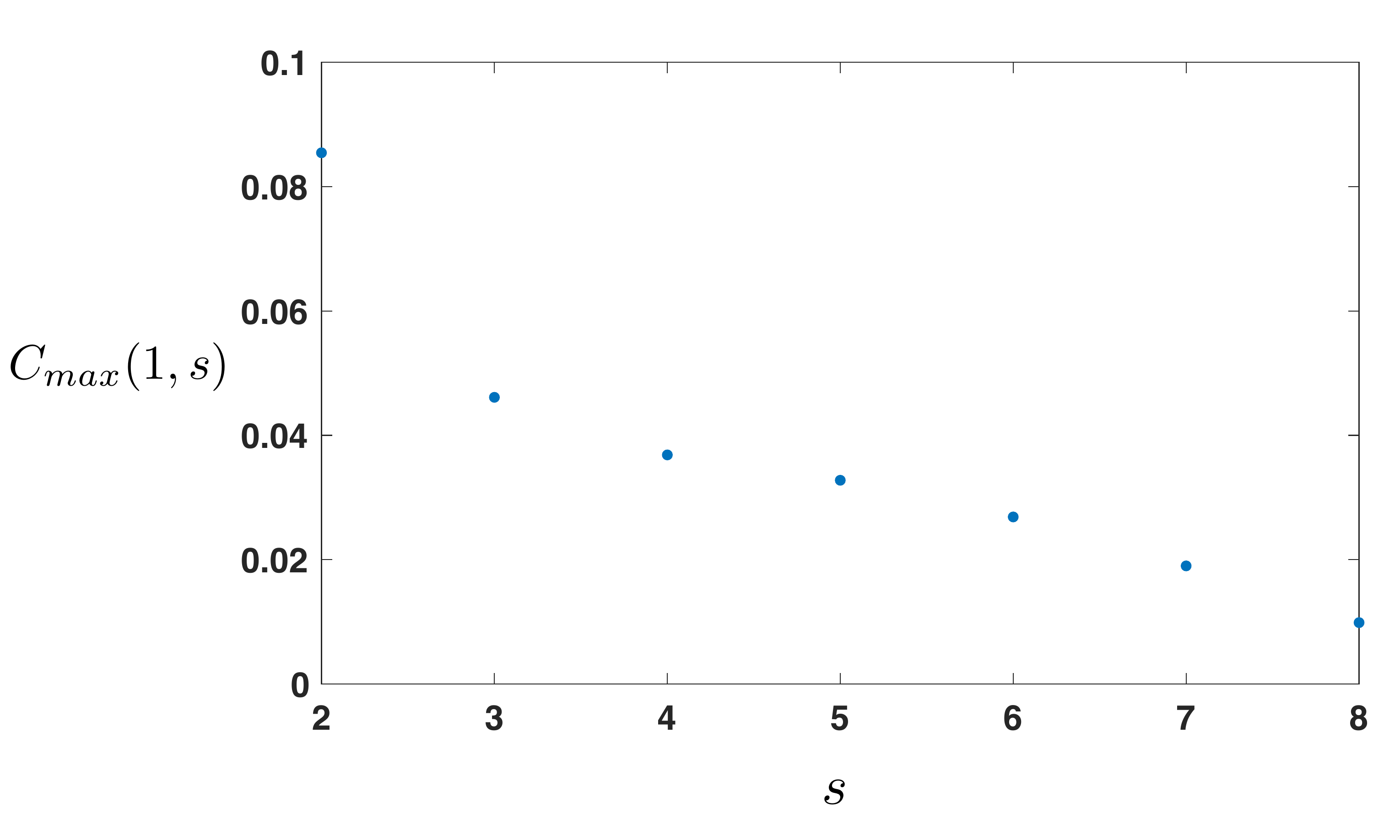}
\caption{Maximum achievable entanglement between sites 1 and $s=2,3,\cdots 8$ by varying $T_R$. Here $N=8, \lambda=1, T_L=0, \Delta=15$ and $g=7.8$ is close to its upper bound in Eq.~(\ref{assump}).}
\label{Fig:Cmax}
\end{figure}

Expected features of the concurrence $C(r,s)$ are that by increasing the distance $s-r$ between the spins with fixed $r$, the maximum achievable bipartite entanglement $C_{max}(r,s)$ diminishes, as shown in Figure \ref{Fig:Cmax} for $C_{max}(1,s)$ in a chain of size $N=8$, \sout{with $T_L=0$, $\lambda=1$, $\Delta=50$ and $g=7.8$ is very close to upper bound in Eq.~(\ref{assump})}  while the concurrence itself vanishes at lower temperatures, in agreement with the fact that distance and temperature play against correlations. Furthermore, the lack of translational invariance makes $C(r,s)$ depend not only on $s-r$, but also on the position $r$ of the first spin.

\begin{figure}[h!]
\centering
\includegraphics[width=\linewidth]{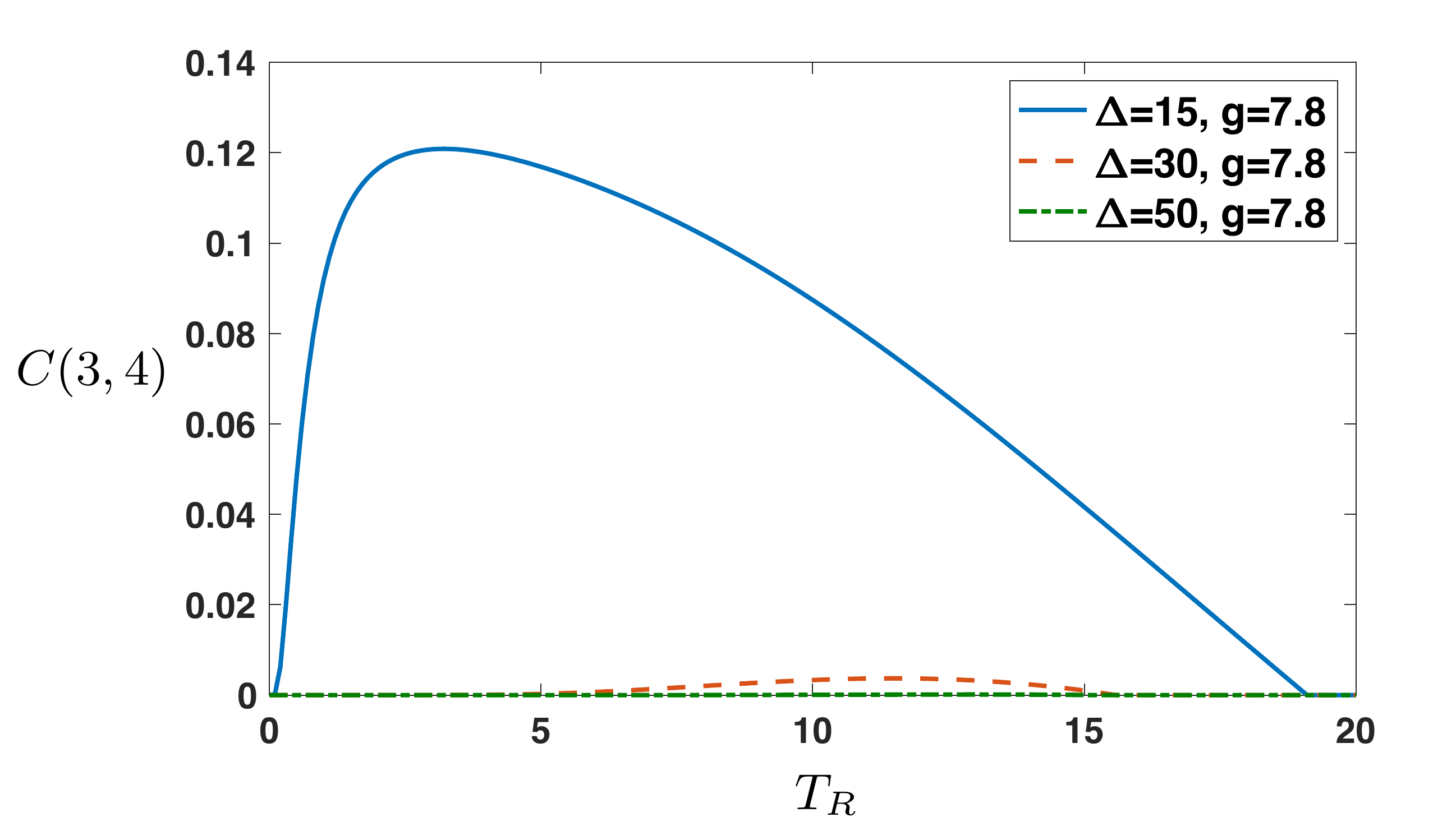}
\caption{Bipartite entanglement between spins $3$ and $4$, as measured by the concurrence $C(3,4)$, versus $T_R$ for $N=8$, $T_L=0$, with
$\Delta=15,30,50$ and $g$ close to the saturation value relative to  $\Delta=15$.}
\label{Fig6:CDG8}
\end{figure}

\begin{figure}[h!]
\centering
\includegraphics[width=\linewidth]{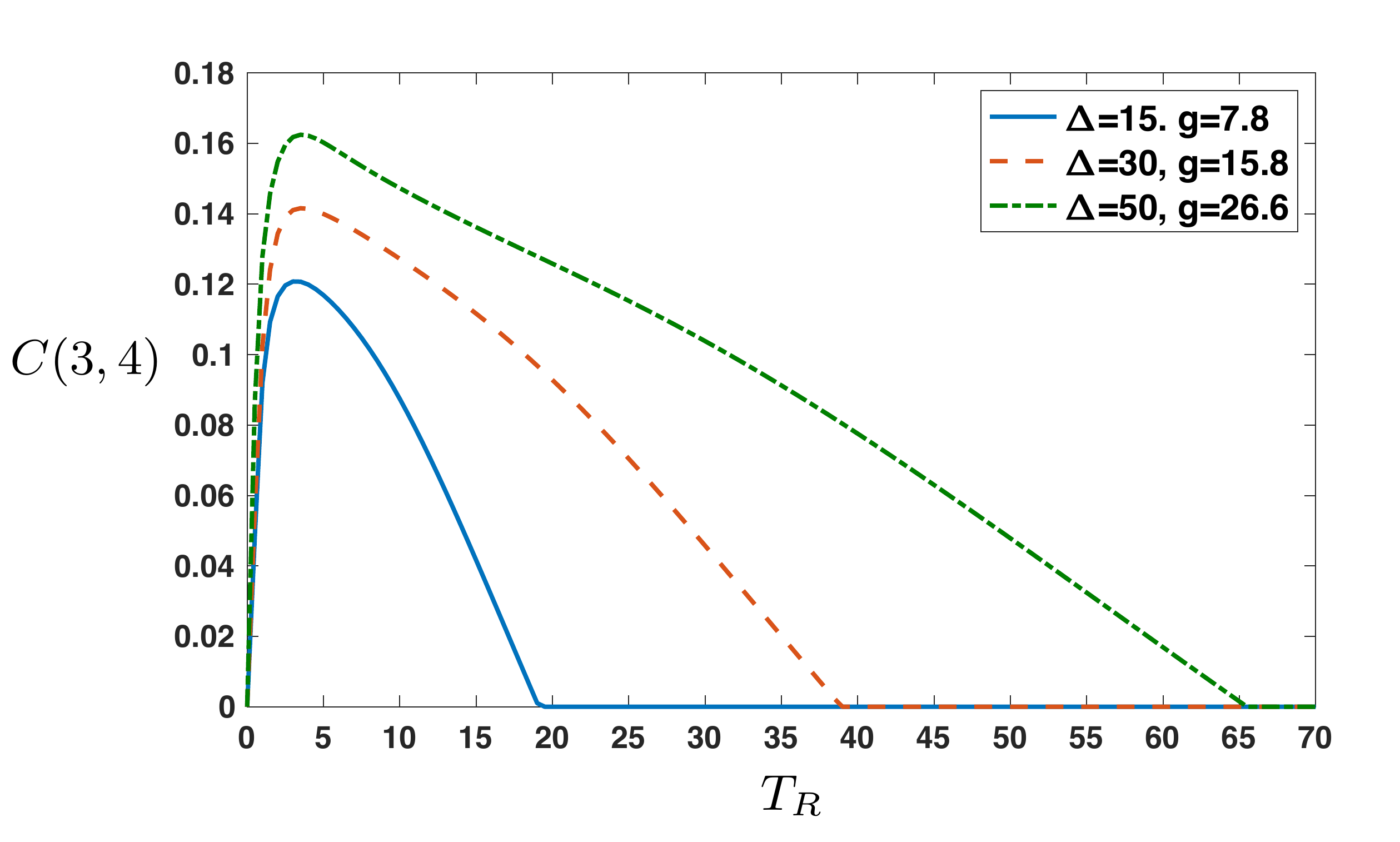}
\caption{Bipartite entanglement between spins $3$ and $4$, as measured by the concurrence $C(3,4)$,
versus $T_R$ for $N=8$, $T_L=0$, with
$\Delta=15,30,50$ and $g$ close to the corresponding saturation values.}
\label{Fig7:CDGO8}
\end{figure}

As regards the dependence of the concurrence on the parameters $\Delta$ and $g$, Figure~\ref{Fig6:CDG8} first shows that, with temperature $T_L=0$, and $g$ fixed, close to the saturation the bound~\eqref{assump} for $\Delta=15$, the entanglement as a function of $T_R$ diminishes while increasing $\Delta$.
%
%
%
This behaviour agrees with the fact that augmenting the transverse external field the spins tend to become all parallel and thus the stationary state separable. On the other hand, by increasing $g$ the spins interact more strongly thus favouring the generation of quantum correlations that may persist asymptotically against temperature.
In fact, the farther is $g$ away from the saturation value at given $\Delta$, the smaller is the achieved entanglement, the larger is the temperature at which it appears and the smaller the one at which it disappears. 
Specifically, \sout{Indeed,} the chosen value of $g$ is sufficient to generate entanglement for $\Delta=15, 30$, but not for $\Delta=50$, while increasing $g$ beyond the saturation value for $\Delta=15$ would violate the condition assumed throughout the manuscript that the transition frequencies $\omega_\ell$ in~\eqref{Lindop9} be positive.
%
%

Instead, in Figure~\ref{Fig7:CDGO8}, under the same conditions as in Figure~\ref{Fig6:CDG8}, the values of the interaction strength $g$ are chosen close to the saturation bound~\eqref{assump} for each $\Delta$. The graph shows that in this case the highest possible $g$, despite the higher values of $\Delta$, contributes to the creation of entanglement as soon as $T_R>0$; moreover, it also makes it last up to higher values of $T_R$.


%
%

\section{Discussion}

Spin chains coupled to external baths at their endpoints represent paradigmatic models for the study
of transport properties in quantum many-body systems, as they allow the precise analysis of the behaviour of spin
and heat flows along the chain. So far, analytic treatments of the asymptotic transport properties in these systems
have been obtained assuming only {\it ad hoc} couplings between the system and the external baths, those that allow
expressing the chain steady state in terms of the so-called ``matrix product states'' and the like \cite{Prosen5}.

Here instead, taking an arbitrary, energy preserving coupling to the end baths, 
we have been able to derive an exact analytic expression for the unique steady state of a generic $N$-sites spin-1/2 chain,
with $XX$-type inter-spin interaction, in a transverse constant magnetic field. 
This has allowed discussing in detail
the open transport properties of the model, treated in the so-called global approach, revealing
the presence of sink and source terms in the spin flow continuity equation, never pointed out before (except in
the $N=3$ case \cite{BFM}).

In addition, having the explicit form of the system asymptotic stationary state allowed analyzing the entanglement properties of the chain. In particular, a procedure has been devised able to algoritmically provide the explicit expression of the reduced two-spin density matrix for any two sites along the chain.
The behaviour of the corresponding entanglement content of the reduced state, as measured by concurrence,  
has been discussed in some relevant cases in terms of the parameters
of the system Hamiltonian and of the bath temperatures.
While increasing bath temperatures and magnitude of the external magnetic field counteracts entanglement, sufficiently high values of the inter-spin interaction coupling constant would always allow the presence of asymptotic entanglement among any couple of  close enough sites.

In addition, these results show that, for generic $N$, there is no apparent relation between the behaviour of heat flow and two-site entanglement as a function of the bath temperatures, as claimed in the literature for the special case $N=2$ \cite{Khandelwal}: entanglement in the chain is generated independently from the heat flow and even in absence of it, as in the case of isothermal baths.
Furthermore, the two quantities behave rather differently with respect to the length of the chain: while the concurrence vanishes as $1/N$, the heat flow does not.

We are confident that these findings will stimulate further research on the use of many-body systems,  and spin-chains in particular, for modelling quantum transport processes, in view of possible applications in quantum techonology.


\appendix

\section{Diagonalization of the spin-chain Hamiltonian}
\label{App1}

Let us consider the Hamiltonian
$$
\widetilde{H}=\gamma\,\sum_{j=1}^Na^\dag_ja_j\,+\,\sum_{j=1}^{N-1}\Big(a^\dag_j\,a_{j+1}\,+\,a^\dag_{j+1}a_j\Big)
$$
and recast it as $\widetilde{H}=\sum_{j,k=1}^Nh_{jk}\,a^\dag_j\,a_k$ where
\begin{equation}
\label{N-Ham5}
h:=\Big[h_{jk}\Big]=\begin{pmatrix}
\gamma&1&0&0&\ldots&0&0\cr
1&\gamma&1&0&\ldots&0&0\cr
0&1&\gamma&1&\ldots&0&0\cr
\dot&\dot&\dot&\dot&\ldots&\dot&\dot\cr
\dot&\dot&\dot&\dot&\ldots&\dot&\dot\cr
\dot&\dot&\dot&\dot&\ldots&\dot&\dot\cr
0&0&0&0&\ldots&\gamma&1\cr
0&0&0&0&\ldots&1&\gamma
\end{pmatrix}=\gamma\,\mathbb{I}\,+\,B\ .
\end{equation}  
The rank $N$ tridiagonal matrix 
\begin{equation}
\label{appA0}
B=\begin{pmatrix}
0&1&0&0&\ldots&0&0\cr
1&0&1&0&\ldots&0&0\cr
0&1&0&1&\ldots&0&0\cr
\dot&\dot&\dot&\dot&\ldots&\dot&\dot\cr
\dot&\dot&\dot&\dot&\ldots&\dot&\dot\cr
\dot&\dot&\dot&\dot&\ldots&\dot&\dot\cr
0&0&0&0&\ldots&0&1\cr
0&0&0&0&\ldots&1&0
\end{pmatrix}\ ,
\end{equation}
can be diagonalized as follows. We shall emphasize the rank by writing $B_N$ instead of $B$. Then, one notices that
\begin{equation}
\label{appA1}
\hbox{det}\Big(x-B_N\Big)=x\,\hbox{det}\Big(x-B_{N-1}\Big)-\hbox{det}\Big(x-B_{N-2}\Big)\ ,
\end{equation}
whence the same equation is satisfied by the associated characteristic polynomial $p_N(x)$:
\begin{equation}
\label{appA2}
p_{N+2}(x)\,-\,x\,p_{N+1}(x)\,+\,p_N(x)=0\ .
\end{equation}
Setting $x=2\cos\theta$, one finds the solution
\begin{equation}
\label{appA3}
p_N(x)=C_1\,{\rm e}^{i\,N\,\theta}\,+\,C_2\,{\rm e}^{-i\,N\,\theta}\ .
\end{equation}
From $p_1(x)=x$ and $p_2(x)=x^2-1$, one fixes the coefficients
\begin{equation}
\label{appA4} 
C_1=\frac{{\rm e}^{i\theta}}{{\rm e}^{i\theta}\,-\,{\rm e}^{-i\theta}}\ ,\qquad C_2=-\frac{{\rm e}^{-i\theta}}{{\rm e}^{i\theta}\,-\,{\rm e}^{-i\theta}}\ ,
\end{equation}
whence
$\displaystyle p_N(x)=\frac{\sin((N+1)\theta)}{\sin\theta}$. 

Since for $\theta=0$, respectively $\theta=\pi$, $p_N(2)=N+1$, respectively $p_N(-2)=(-)^N(N+1)$, the only zeroes of $p_N(x)$ are at
$\displaystyle\theta_k=\frac{k\,\pi}{N+1}$, $k=1,2,\ldots,N$. It thus follows that the eigenvalues of $h=\gamma\,\mathbb{I}+B$ are
\begin{equation}
\label{appA6} 
\lambda_k=\gamma\,+\,2\,\cos\left(\frac{k\,\pi}{N+1}\right)\ .
\end{equation}
Finally, one can check that the symmetric matrix 
\begin{equation}
\label{appA7} 
U=\Big[u_{jk}\Big]_{j,k=1}^N\ ,\ u_{jk}=\sqrt{\frac{2}{N+1}}\,\sin\left(\frac{j\,k\,\pi}{N+1}\right)\ .
\end{equation}
is also orthogonal. It can be directly checked that
\begin{equation}
\label{appA8} 
U\,B\,U={\rm diag}\left[2\,\cos\left(\frac{\ell\pi}{N+1}\right)\right]_{\ell=1}^N\ .
\end{equation}

\section{Energies and eigenvectors of two and three spin chains}
\label{App_2_3:sec}

\subsection{Two-spin chain}
\label{2spinsp}

For a chain consisting of two spins only as in~\cite{Levy}--~\cite{Haack2}, $N=2$ and~\eqref{N_Ham11} yield
the energies, 
\begin{equation}
\label{2spinen}
E_{00}=-2\,\Delta,\ E_{10}=2\,g,\ E_{01}=-2\,g,\ E_{11}=2\,\Delta\ .
\end{equation}
Furthermore, from~\eqref{appA7}, the unitary matrix $U$ results
\begin{equation}
\label{2spinunmat}
U=\frac{1}{\sqrt{2}}\begin{pmatrix}1&1\cr 1&-1\end{pmatrix}\ .
\end{equation}
Therefore, applying~\eqref{JW5a} one obtains
\begin{align}
\label{2spinbop1}
b_1&=\frac{\sigma^{(1)}_--\sigma^{(1)}_z\sigma^{(2)}_-}{\sqrt{2}}\ ,\ b_2=\frac{\sigma^{(1)}_-+\sigma^{(1)}_z\sigma^{(2)}_-}{\sqrt{2}}\ ,\\
\label{2spinbop2}
b_1b_2&=-\sigma^{(1)}_-\sigma^{(2)}_-\ ,
\end{align}
so that using~\eqref{N_Ham10} and~\eqref{vac1},
one can recast the eigenvectors $\vert n_1n_2\rangle$ relative to the eigenvalues $E_{n_1n_2}$ using the standard basis,  $\vert\uparrow\rangle$,
$\vert\downarrow\rangle$. 
Indeed, 
$\vert vac\rangle=\vert 00\rangle=\vert \downarrow\downarrow\rangle$, whence
\begin{align}
\label{2spineigenv1}
\vert 00\rangle&=\vert\downarrow\downarrow\rangle\ , & \vert10\rangle&=\frac{\vert\uparrow\downarrow\rangle+\vert\downarrow\uparrow\rangle}{\sqrt{2}}\\
\label{2spineigenv2}
\vert01\rangle&=\frac{\vert\uparrow\downarrow\rangle-\vert\downarrow\uparrow\rangle}{\sqrt{2}}\ , & \vert 11\rangle&=\vert\uparrow\uparrow\rangle\ .
\end{align}

\subsection{Three spin chain}
\label{3spisp}

In the case of a three-spin chains~\cite{BFM}, setting $N=3$  the eigenvalues of the Hamiltonian~\eqref{spin-hamiltonian} are 
\begin{equation}
\label{oldeigval}
E_{n_1n_2n_3}=\Delta\,\Big(2\,\sum_{\ell=1}^3\,n_\ell-3\Big)+4g\sum_{\ell=1}^3n_\ell\,\cos\left(\frac{\ell\pi}{4}\right)\ .
\end{equation}
Explicitly they and their corresponding eigenvectors read
\begin{align}
\label{oldeigval1}
\vert 000\rangle&=\vert vac\rangle\ ,\ &E_{000}&=-3\Delta\ ,\\
\label{oldeigval2}
\vert 100\rangle&=b^\dag_1\vert000\rangle\ ,\ &E_{100}&=-\,\Delta\,+\,2\sqrt{2}g\ ,\\
\label{oldeigval3}
\vert 010\rangle&=b^\dag_2\vert000\rangle\ ,\ &E_{010}&=-\,\Delta\ ,\\
\label{oldeigval4}
\vert 001\rangle&=b^\dag_3\vert000\rangle\ , &E_{001}&=-\,\Delta\,-\,2\sqrt{2}g\ ,\\
\label{oldeigval5}
\vert 110\rangle&=b^\dag_1b^\dag_2\vert000\rangle\ , &E_{110}&=\Delta\,+\,2\sqrt{2}g\ ,\\
\label{oldeigval6}
\vert 101\rangle&=b^\dag_1b^\dag_3\vert000\rangle\ , &E_{101}&=\Delta\ ,\\
\label{oldeigval7}
\vert 011\rangle&=b^\dag_2b^\dag_3\vert000\rangle\ , &E_{011}&=\Delta\,-\,2\sqrt{2}g\,\\
\label{oldeigval8}
\vert 111\rangle&=b^\dag_1b^\dag_2b^\dag_3\vert000\rangle\ , &E_{111}&=3\,\Delta\ .
\end{align}
The correspondence with the eigenvalues $E_j$ in~\cite{BFM} is as follows
\begin{eqnarray*}
&&
E_{000}=E_2\ ,\ E_{100}=E_7\ ,\ E_{010}=E_4\ ,\ E_{001}=E_6\ ,\\
&&
E_{110}=E_5\ ,\ E_{101}=E_3\ ,\ E_{011}=E_8\ ,\ E_{111}=E_1\ .
\end{eqnarray*}
Since the $3\times 3$ matrix $U$ in~\eqref{appA7} reads
\begin{equation}
\label{3mat1}
U=\frac{1}{2}\begin{pmatrix}
1&\sqrt{2}&1\cr
\sqrt{2}&0&-\sqrt{2}\cr
1&-\sqrt{2}&1
\end{pmatrix}\ .
\end{equation}
Then, $\sigma_\pm\sigma_z=\mp\sigma_\pm$ and \eqref{JW5} yield
\begin{align}
\label{opb1}
b^\dag_1&=\frac{1}{2}\Big(\sigma^{(1)}_+\,-\,\sqrt{2}\,\sigma^{(1)}_z\,\sigma^{(2)}_+\,+\,\sigma^{(1)}_z\,\sigma^{(2)}_z\,\sigma^{(3)}_+\Big)\\
\label{opb2}
b^\dag_2&=\frac{1}{\sqrt{2}}\Big(\sigma^{(1)}_+\,-\,\sigma^{(1)}_z\,\sigma^{(2)}_z\,\sigma^{(3)}_+\Big)\\ 
\label{opb3}
b^\dag_3&=\frac{1}{2}\Big(\sigma^{(1)}_+\,+\,\sqrt{2}\,\sigma^{(1)}_z\,\sigma^{(2)}_+\,+\,\sigma^{(1)}_z\,\sigma^{(2)}_z\,\sigma^{(3)}_+\Big)\\
\label{opb12}
b^\dag_1b^\dag_2&=-\frac{1}{2}\Big(\sigma^{(1)}_+\,\sigma^{(2)}_+\,-\,\sqrt{2}\,\sigma^{(1)}_+\,\sigma^{(2)}_z\,\sigma^{(3)}_+\,+\,\sigma^{(2)}_+\,\sigma^{(3)}_+\Big)\\
\label{opb23}
b^\dag_2b^\dag_3&=-\frac{1}{2}\Big(\sigma^{(1)}_+\,\sigma^{(2)}_+\,+\,\sqrt{2}\,\sigma^{(1)}_+\,\sigma^{(2)}_z\,\sigma^{(3)}_+\,+\,\sigma^{(2)}_+\,\sigma^{(3)}_+\Big)\\
\label{opb13}
b^\dag_1b^\dag_3&=-\frac{1}{\sqrt{2}}\Big(\sigma^{(1)}_+\,\sigma^{(2)}_+\,-\,\sigma^{(2)}_+\,\sigma^{(3)}_+\Big)\\
\label{opb123}
b^\dag_1b^\dag_2 b^\dag_3&=-\sigma^{(1)}_+\,\sigma^{(2)}_+\,\sigma^{(3)}_+\ .
\end{align}
The expressions of  the eigenvectors $\vert n_1n_2n_3\rangle$ in the spin standard basis  and  their correspondence with the eigenvectors 
obtained in~\cite{BFM} are reported in Appendix~\ref{Appaid}.

\section{Lindblad operators for two and three spin chains}
\label{app_2_3_open:sec}
\subsection{Two-spin chain}
\label{2spin_bath}

In the case of $N=2$, from~\eqref{Lindop9}, one computes the following  transition frequencies
\begin{equation}
\label{2spinfreq2}
\omega_1=2(\Delta+g)\ ,\ \omega_2=2(\Delta-g)\ .
\end{equation}
Using~\eqref{Lindop10} and~\eqref{Lindop11b},
the following Lindblad operators ensue for the open two-spin chain:
\begin{align}
\label{2spinLindop1}
A_L^\dag(\omega_1)&=\frac{\vert 10\rangle]\langle00\vert+\vert11\rangle\langle01\vert}{\sqrt{2}}\ ,\\
\label{2spinLindop2}
A_L^\dag(\omega_2)&=\frac{\vert 01\rangle]\langle00\vert-\vert11\rangle\langle10\vert}{\sqrt{2}}\ ,\\
\label{2spinLindop3}
A_R^\dag(\omega_1)&=\frac{\vert 10\rangle]\langle00\vert-\vert11\rangle\langle0\vert}{\sqrt{2}}\ ,\\
\label{2spinLindop3}
A_R^\dag(\omega_2)&=-\frac{\vert 01\rangle]\langle00\vert+\vert11\rangle\langle10\vert}{\sqrt{2}}\ .
\end{align}
According to the discussion before Remark~\ref{rem-imp}, in order to have all of them contribute to dissipation, one must set
$g\leq \Delta$.

\subsection{Three-spin chain}
\label{Appaid}

Setting $N=3$, from~\eqref{Lindop9} 
we get the three frequencies
\begin{equation}
\label{3s1}
\omega_1=2\Big(\Delta+\sqrt{2}g\Big),\ \omega_2=2\Delta,\ \omega_3=2\Big(\Delta-\sqrt{2}g\Big)\ .
\end{equation}
They correspond to the three frequencies $\omega_1$, $\omega_0$ and $\omega_2$ in~\cite{BFM}.
Furthermore,~\eqref{Lindop10} and~\eqref{Lindop11b} yield the left-bath Lindblad operators
\begin{align}
\nonumber
A^\dag_L(\omega_1)&=\frac{1}{2}\Big(\vert 100\rangle\langle000\vert\,+\,\vert110\rangle\langle010\vert\,+\,\vert101\rangle\langle001\vert\\
\label{3sL1}
&\hskip 1cm
+\,\vert111\rangle\langle011\vert\Big)\\
\nonumber
A^\dag_L(\omega_2)&=\frac{1}{\sqrt{2}}\Big(\vert 010\rangle\langle000\vert\,-\,\vert110\rangle\langle100\vert\,-\,\vert111\rangle\langle101\vert\\
\label{3sL2}
&\hskip 1cm
+\,\vert011\rangle\langle001\vert\Big)\\
\nonumber
A^\dag_L(\omega_3)&=\frac{1}{2}\Big(\vert 001\rangle\langle000\vert\,-\,\vert101\rangle\langle100\vert\,-\,\vert011\rangle\langle010\vert\\
\label{3sL3}
&\hskip 1cm
+\,\vert111\rangle\langle110\vert\Big)\ ,
\end{align}
and the right bath Lindblad operators
\begin{align}
\nonumber
A^\dag_R(\omega_1)&=\frac{1}{2}\Big(\vert011\rangle\langle001\vert\,+\,\vert111\rangle\langle101\vert\,-\,\vert 010\rangle\langle000\vert\\
\label{3sR1}
&\hskip 1cm
-\,\vert110\rangle\langle100\vert\Big)\\
\nonumber
A^\dag_R(\omega_2)&=\frac{1}{\sqrt{2}}\Big(\vert011\rangle\langle001\vert\,
+\,\vert111\rangle\langle101\vert\,-\,\vert 010\rangle\langle000\vert\\
\label{3sR2}
&\hskip 1cm
-\,\vert110\rangle\langle100\vert\Big)\\
\nonumber
A^\dag_R(\omega_3)&=\frac{1}{2}\Big(\vert 001\rangle\langle000\vert\,+\,\vert101\rangle\langle100\vert\,+\,\vert011\rangle\langle010\vert\\
\label{3sR3}
&\hskip 1cm
+\,\vert111\rangle\langle110\vert\Big)\ .
\end{align}
According to the discussion before Remark~\ref{rem-imp}, in order to have all of them contribute to dissipation, one must set
$\displaystyle g\leq \frac{\Delta}{\sqrt{2}}$.

\label{Appaid}


In terms of the  eigenstates $\vert\uparrow\rangle$ and $\vert\downarrow\rangle$ of $\sigma_z$, using that $\vert vac\rangle=\vert 000\rangle=\vert\downarrow\downarrow\downarrow\rangle$, one can reexpress the eigenvectors $\vert n_1n_2n_3\rangle$ 
in the spin standard basis:  
\begin{equation}
\label{oldeigvec1}
\vert 000\rangle=\vert\downarrow\downarrow\downarrow\rangle=\vert E_2\rangle\ ,
\end{equation}
in the case of zero excitations, while for one excitation,
\begin{align}
\nonumber
\vert 100\rangle&=b^\dag_1\vert000\rangle=\frac{1}{2}
\Big(\vert\uparrow\downarrow\downarrow\rangle\,+\,\sqrt{2}\,\vert\downarrow\uparrow\downarrow\rangle\,+\,
\vert\downarrow\downarrow\uparrow\rangle\Big)\ ,\\
\label{oldeigvec2}
&=\vert E_7\rangle\\
\label{oldeigvec3}
\vert 010\rangle&=b^\dag_2\vert000\rangle=\frac{1}{\sqrt{2}}
\Big(\vert\uparrow\downarrow\downarrow\rangle\,-\,\vert\downarrow\downarrow\uparrow\rangle\Big)
=\vert E_4\rangle\ ,\\
\nonumber
\vert 001\rangle&=b^\dag_3\vert000\rangle=\frac{1}{2}
\Big(\vert\uparrow\downarrow\downarrow\rangle\,-\,\sqrt{2}\,\vert\downarrow\uparrow\downarrow\rangle\,+\,
\vert\downarrow\downarrow\uparrow\rangle\Big)\\
\label{oldeigvec2}
&=\vert E_6\rangle\ ,
\end{align}
and for two excitations 
\begin{align}
\nonumber
\vert 110\rangle&=b^\dag_1b^\dag_2\vert000\rangle=-\frac{1}{2}
\Big(\vert\uparrow\uparrow\downarrow\rangle\,+\,\sqrt{2}\,\vert\uparrow\downarrow\uparrow\rangle\,+\,
\vert\downarrow\uparrow\uparrow\rangle\Big)\ ,\\
\label{oldeigvec5}
&=-\vert E_5\rangle\ ,\\
\nonumber
\vert 101\rangle&=b^\dag_1b^\dag_3\vert000\rangle=-\frac{1}{\sqrt{2}}
\Big(\vert\uparrow\uparrow\downarrow\rangle\,-\,\vert\downarrow\uparrow\uparrow\rangle\Big)\ ,\\
\label{oldeigvec6}
&=-\vert E_3\rangle\\
\nonumber
\vert 011\rangle&=b^\dag_2b^\dag_3\vert000\rangle=-\frac{1}{2}
\Big(\vert\uparrow\uparrow\downarrow\rangle\,-\,\sqrt{2}\,\vert\uparrow\downarrow\uparrow\rangle\,+\,
\vert\downarrow\uparrow\uparrow\rangle\Big)\\
\label{oldeigvec7}
&=-\vert E_8\rangle\ .
\end{align}
Finally, in the case of three excitations one finds
\begin{equation}
\label{oldeigvec8}
\vert 111\rangle=b^\dag_1b^\dag_2b^\dag_3\vert000\rangle=-\vert\uparrow\uparrow\uparrow\rangle =-\vert E_1\rangle\ ,
\end{equation}
the difference in the overall sign depending on the chosen ordering of the creation operators $b^\dag_\ell$.
The correspondence with the $\vert E_k\rangle$
obtained in~\cite{BFM} is confirmed after noticing that there $\vert 0\rangle=\vert\uparrow\rangle$ and $\vert 1\rangle=\vert\downarrow\rangle$.

In the same vain, using~\eqref{oldeigvec1}--~\eqref{oldeigvec8}, one recasts the Lindblad operators in~\eqref{3sL1}--~\eqref{3sR3} as
\begin{align}
\nonumber
A^\dag_L(\omega_1)&=\frac{1}{2}\Big(\vert E_7\rangle\langle E_2\vert\,-\,\vert E_5\rangle\langle E_4\vert\,-\,\vert E_3\rangle\langle E_6\vert\\
\label{3sLold1}
&\hskip 1cm
+\,\vert E_1\rangle\langle E_8\vert\Big)\ ,\\
\nonumber
A^\dag_L(\omega_2)&=\frac{1}{\sqrt{2}}\Big(\vert E_4\rangle\langle E_2\vert\,+\,\vert E_5\rangle\langle E_7\vert\,-\,\vert E_1\rangle\langle E_3\vert\\
\label{3sLold0}
&\hskip 1cm
-\,\vert E_8\rangle\langle E_6\vert\Big)\ ,\\
\nonumber
A^\dag_L(\omega_3)&=\frac{1}{2}\Big(\vert E_6\rangle\langle E_2\vert\,+\,\vert E_3\rangle\langle E_7\vert\,+\,\vert E_8\rangle\langle E_4\vert\\
\label{3sLold2}
&\hskip 1cm
+\,\vert E_1\rangle\langle E_5\vert\Big)\ ,
\end{align}
and
\begin{align}
\nonumber
A^\dag_R(\omega_1)&=\frac{1}{2}\Big(\vert E_7\rangle\langle E_2\vert\,+\,\vert E_5\rangle\langle E_4\vert\,+\,\vert E_3\rangle\langle E_6\vert\\
\label{3sRold1}
&\hskip 1cm
+\,\vert E_1\rangle\langle E_8\vert\Big)\\
\nonumber
A^\dag_R(\omega_2)&=\frac{1}{\sqrt{2}}\Big(\vert E_5\rangle\langle E_7\vert\,-\,\vert E_4\rangle\langle E_2\vert\,-\,\vert E_8\rangle\langle E_6\vert\\
\label{3sRold0}
&\hskip 1cm
+\,\vert E_1\rangle\langle E_3\vert\Big)\\
\nonumber
A^\dag_R(\omega_3)&=\frac{1}{2}\Big(\vert E_6\rangle\langle E_2\vert\,+\,\vert E_1\rangle\langle E_5\vert\,-\,\vert E_3\rangle\langle E_7\vert\\
\label{3sRold2}
&\hskip 1cm
-\,\vert E_8\rangle\langle E_4\vert\Big)\ .
\end{align}
These expressions coincide with those found in~\cite{BFM}.

\section{Uniqueness of the stationary state}
\label{App3}

Consider the commutator of a generic chain operator $X=\sum_{\mathbf{p},\mathbf{q}}\,X_{\mathbf{p}\mathbf{q}}\,\vert\mathbf{p}\rangle\langle\mathbf{q}\vert$,  
in the energy eigenbasis~\eqref{Lindop3c}, with the Lindblad operator in~\eqref{Lindop10} and set it equal to zero:
\begin{eqnarray*}
&&
\langle\mathbf{r}\vert\left[X\,,\,A^\dag_L(\omega_\ell)\right]\vert\mathbf{s}\rangle
=u_{1\ell}\,\sum_{\mathbf{p},\mathbf{q},\widehat{\mathbf{n}}_\ell}(-1)^{\sum_{j=1}^{\ell-1}n_j}\,X_{\mathbf{p}\mathbf{q}}\,\times\\
&&\hskip 1cm
\times\langle\mathbf{r}\vert\Big[\vert\mathbf{p}\rangle\langle\mathbf{q}\vert\,,\,\vert \mathbf{n}_{1_\ell}\rangle\langle \mathbf{n}_{0_\ell}\vert\Big]\vert\mathbf{s}\rangle=0\ .
\end{eqnarray*}
This yields
\begin{equation}
\label{app3.1}
X_{\mathbf{r}\mathbf{s}_{1_\ell}}(-1)^{\sum_{j=1}^{\ell-1}s_j}\,\delta_{s_\ell0}\,=\,X_{\mathbf{r}_{0_\ell}\mathbf{s}}(-1)^{\sum_{j=1}^{\ell-1}r_j}\,\delta_{r_\ell1}\ ,
\end{equation}
whence, choosing $r_\ell=0$ and $s_\ell=0$ gives $X_{\mathbf{r}_{0_\ell}\mathbf{s}_{1_\ell}}=0$.
Changing $A^\dag_L(\omega_\ell)$ into $A^\dag_R(\omega_\ell)$ yields $X_{\mathbf{r}_{1_\ell}\mathbf{s}_{0_\ell}}=0$; thus, the only 
non-vanishing $X$ commuting with all Lindblad operators must be diagonal in the energy eigenbasis: namely,
 $X=\sum_{\mathbf{n}}\,X_{\mathbf{n}\mathbf{n}}\vert\mathbf{n}\rangle\langle\mathbf{n}\vert$.
On the other hand, choosing $r_\ell=1$ and $s_\ell=0$,~\eqref{app3.1} yields $X_{\mathbf{n}_{1_\ell}\mathbf{n}_{1_\ell}}=X_{\mathbf{n}_{0_\ell}\mathbf{n}_{0_\ell}}$ for all $\ell=1,2,\ldots,N$, whence $X$ must be a multiple of the identity.

\section{Sink and source contributions to  the stationary transport properties}
\label{App5}

Given the stationary state in~\eqref{SS12}, in order to compute 
\begin{equation}
\label{App5.1}
\mathfrak{Q}^{(k)}_L=\lambda^2\sum_{\ell=1}^N\text{Tr}\big(\rho_\infty\,\widetilde{\mathbb{D}}^{(L)}_{\omega_\ell}[\sigma^{(k)}_z]\big)\ ,
\end{equation}
in~\eqref{sinkssources1}, we need evaluate mean-values of the form
\begin{eqnarray}
\nonumber
&&
\langle\mathbf{n}\vert\widetilde{\mathbb{D}}^{(L)}_{\omega_\ell}[\sigma^{(k)}_z]\vert\mathbf{n}\rangle=
C^{(L)}_{\omega_\ell}\,\left(\langle\mathbf{n}\vert A^\dag_L(\omega_\ell)\sigma^{(k)}_z A_L(\omega_\ell)\vert\mathbf{n}\rangle\right.\\
\label{App5.2a}
&&
\hskip 1cm
\left.-\frac{1}{2}\langle\mathbf{n}\vert\left\{A^\dag_L(\omega_\ell)A_L(\omega_\ell),\sigma^{(k)}_z\right\}\vert\mathbf{n}\rangle\right)\\
\nonumber
&&\hskip 2.5cm
+\widetilde{C}^{(L)}_{\omega_\ell}\,\left(\langle\mathbf{n}\vert A_L(\omega_\ell)\sigma^{(k)}_z A^\dag_L(\omega_\ell)\vert\mathbf{n}\rangle\right.\\
\label{App5.2b}
&&
\hskip 1cm
\left.-\frac{1}{2}\langle\mathbf{n}\vert\left\{A_L(\omega_\ell)A^\dag_L(\omega_\ell),\sigma^{(k)}_z\right\}\vert\mathbf{n}\rangle\right)\ .
\end{eqnarray}
Using the expressions~\eqref{Lindop10},~\eqref{SS1a} and~\eqref{SS1c} one gets
\begin{eqnarray}
\label{App5.3a}
&&
\hskip-.6cm
\langle\mathbf{n}\vert A^\dag_L(\omega_\ell)\sigma^{(k)}_z A_L(\omega_\ell)\vert\mathbf{n}\rangle=u_{1\ell}^2\,\delta_{1n_\ell}\,\langle\mathbf{n}_{0_\ell}\vert\sigma^{(k)}_z\vert\mathbf{n}_{0_\ell}\rangle,\\
\label{App5.3b}
&&
\hskip-.6cm
\langle\mathbf{n}\vert\left\{A^\dag_L(\omega_\ell)A_L(\omega_\ell),\sigma^{(k)}_z\right\}\vert\mathbf{n}\rangle=u_{1\ell}^2\,\delta_{1n_\ell}\,\langle\mathbf{n}\vert\sigma^{(k)}_z\vert\mathbf{n}\rangle,\\
\label{App5.3c}
&&
\hskip-.6cm
\langle\mathbf{n}\vert A_L(\omega_\ell)\sigma^{(k)}_z A^\dag_L(\omega_\ell)\vert\mathbf{n}\rangle=u_{1\ell}^2\,\delta_{0n_\ell}\,\langle\mathbf{n}_{1_\ell}\vert\sigma^{(k)}_z\vert\mathbf{n}_{1_\ell}\rangle,\\
\label{App5.3d}
&&
\hskip-.6cm
\langle\mathbf{n}\vert\left\{A_L(\omega_\ell)A^\dag_L(\omega_\ell),\sigma^{(k)}_z\right\}\vert\mathbf{n}\rangle=u_{1\ell}^2\,\delta_{0n_\ell}\,\langle\mathbf{n}\vert\sigma^{(k)}_z\vert\mathbf{n}\rangle.
\end{eqnarray}
Then, from~\eqref{JW3} and~\eqref{JW4} it follows that
\begin{equation}
\label{App5.4}
\sigma^{(k)}_z=-1\,+\,2\,\sum_{r,s=1}^Nu_{kr}\,u_{ks}\,b^\dag_r\,b_s\ ,
\end{equation}
whence, using~\eqref{Lindop3a}--~\eqref{Lindop3c},
\begin{eqnarray}
\label{App5.5a}
\hskip-.5cm
\langle\mathbf{n}\vert\sigma^{(k)}_z\vert\mathbf{n}\rangle&=&2\,\sum_{r=1}^N\,n_r\,u^2_{kr}\,-\,1\ ,
\\
\label{App5.5b}
\hskip-.5cm
\langle\mathbf{n}_{0_\ell}\vert\sigma^{(k)}_z\vert\mathbf{n}_{0_\ell}\rangle&-&\langle\mathbf{n}\vert\sigma^{(k)}_z\vert\mathbf{n}\rangle=
-2\,n_\ell\,u^2_{k\ell}\ ,
\\
\label{App5.5c}
\hskip-.5cm
\langle\mathbf{n}_{1_\ell}\vert\sigma^{(k)}_z\vert\mathbf{n}_{1_\ell}\rangle&-&\langle\mathbf{n}\vert\sigma^{(k)}_z\vert\mathbf{n}\rangle
=2(1-n_\ell)\,u^2_{k\ell}\ .
\end{eqnarray}
Inserting the last three expressions into~\eqref{App5.2a}--~\eqref{App5.3d} yields
\begin{equation}
\label{App5.6}
\hskip-.3cm
\langle\mathbf{n}\vert\widetilde{\mathbb{D}}^{(L)}_{\omega_\ell}[\sigma^{(k)}_z]\vert\mathbf{n}\rangle=
u^2_{k\ell}u^2_{1\ell}\Big(\widetilde{C}^{(L)}_{\omega_\ell}\delta_{0n_\ell}-C^{(L)}_{\omega_\ell}\delta_{1n_\ell}\Big),
\end{equation}
whence, finally using the stationary state eigenvalues in~\eqref{SS12}--~\eqref{SS12e} and the explicit form of the constants $C^{(L)}_{\omega_\ell}$ and
$\widetilde{C}^{(L)}_{\omega_\ell}$  in~\eqref{MS4a},~\eqref{MS4b},  the sink/source contribution $\mathfrak{Q}^{(k)}_L$ in~\eqref{sinkssources1} ensues.
Similar arguments lead to $\mathfrak{Q}^{(k)}_R$ in~\eqref{sinkssources2}.

 \section{Stationary state eigenvalues for two and three spin chains}
\label{app_2_3_stst:sec}

\subsection{Two-spin chain}
\label{2spinstst}

For two spins and $N=2$, the four $2$-digit strings are, in anti-lexicographic order,  
$(00), (10), (01)$ and $(11)$. Moreover, using~\eqref{combenum} their combinadic ordering shows to be the same; namely,
\begin{equation}
\label{2spincombinadic}
(00)\leftrightarrow\mathcal{N}_0=1\ ,\
\left\{\begin{matrix}
(10)\leftrightarrow{\mathcal{N}}_1=1\cr 
(01)\leftrightarrow\mathcal{N}_1=2\end{matrix}\right.\ ,\  (11)\leftrightarrow\mathcal{N}_2=1\ .
\end{equation}
Then, the combinadic list of the eigenvalues $\Lambda_{n_1n_2}$ of the stationary state $\rho_\infty$ in~\eqref{SS12a} are 
\begin{align}
\label{2spststcombrep0}
\mathcal{L}^{(0)}_1&=\Lambda_{00}=\frac{R^{(1)}_0R^{(2)}_0}{R_1R_2}\ ,\ \hbox{for $p=0$} \ ,\\
\label{2spststcombrep11}
\mathcal{L}^{(1)}_1&=\Lambda_{10}=\frac{R^{(1)}_1R^{(2)}_0}{R_1R_2}\ , \
\mathcal{L}^{(1)}_2=\Lambda_{01}=\frac{R^{(1)}_0R^{(2)}_1}{R_1R_2}
\end{align}
for $p=1$ and, for $p=2$, 
\begin{equation}
\label{2spststcombrep21}
\mathcal{L}^{(2)}_1=\Lambda_{11}=\frac{R^{(1)}_1R^{(2)}_1}{R_1R_2}\ .
\end{equation}

\subsection{Three-spin chains}
\label{3spinstst}

For $N=3$, let us consider the  anti-lexicograhic ordering of the eight $3$-digit strings:
$$
(000), (100), (010), (110), (001), (101), (011), (111)\ .
$$
Application of~\eqref{combenum} shows that the previous one and the combinadic ordering coincide, in the sense that
\begin{eqnarray}
\label{combinadic1}
&&
(000)\leftrightarrow\mathcal{N}_0=1\ ,\
\left\{\begin{matrix}
(100)\leftrightarrow{\mathcal{N}}_1=1\cr 
(010)\leftrightarrow\mathcal{N}_1=2\cr
 (001)\leftrightarrow\mathcal{N}_1=3\end{matrix}\right.\ ,\\
\label{combinadic2}
&&
\left\{\begin{matrix}
(110)\leftrightarrow{\mathcal{N}}_2=1\cr
(101)\leftrightarrow\mathcal{N}_2=2\cr
(011)\leftrightarrow\mathcal{N}_2=3
\end{matrix}\right.\ ,\ (111)\leftrightarrow\mathcal{N}_3=1\ .
\end{eqnarray}
Then, for a three spin chain, the eigenvalues $\Lambda_{n_1n_2n_3}$ of the stationary state $\rho_\infty$ in~\eqref{SS12a} are 
\begin{equation}
\label{ststcombrep0}
\mathcal{L}^{(0)}_1=\Lambda_{000}=\frac{R^{(1)}_0R^{(2)}_0R^{(3)}_0}{R^{(1)}R^{(2)}R^{(3)}}\ ,
\end{equation}
for $p=0$, while for $p=1$
\begin{eqnarray}
\label{ststcombrep11}
\mathcal{L}^{(1)}_1&=&\Lambda_{100}=\frac{R^{(1)}_1R^{(2)}_0R^{(3)}_0}{R_1R_2R_3}\ ,\\
\label{ststcombrep12}
\mathcal{L}^{(1)}_2&=&\Lambda_{010}=\frac{R^{(1)}_0R^{(2)}_1R^{(3)}_0}{R_1R_2R_3}\ ,\\
\label{ststcombrep13}
\mathcal{L}^{(1)}_3&=&\Lambda_{001}=\frac{R^{(1)}_0R^{(2)}_0R^{(3)}_1}{R_1R_2R_3}\ ,
\end{eqnarray}
for $p=2$,
\begin{eqnarray}
\label{ststcombrep21}
\mathcal{L}^{(2)}_1&=&\Lambda_{110}=\frac{R^{(1)}_1R^{(2)}_1R^{(3)}_0}{R_1R_2R_3}\ ,\\
\label{ststcombrep22}
\mathcal{L}^{(2)}_2&=&\Lambda_{101}=\frac{R^{(1)}_1R^{(2)}_0R^{(3)}_1}{R_1R_2R_3}\ ,\\
\label{ststcombrep23}
\mathcal{L}^{(2)}_3&=&\Lambda_{011}=\frac{R^{(1)}_0R^{(2)}_1R^{(3)}_1}{R_1R_2R_3}\ ,
\end{eqnarray}
and, finally, for $p=3$,
\begin{equation}
\label{ststcombrep31}
\mathcal{L}^{(3)}_1=\Lambda_{111}=\frac{R^{(1)}_1R^{(2)}_1R^{(3)}_1}{R_1R_2R_3}\ .
\end{equation}
The quantities 
\begin{eqnarray}
\nonumber
R^{(\ell)}_0&=&[h_L(\omega_\ell)]^2\Big(1+n_L(\omega_\ell)\Big)\\
\label{3spinstst0}
&&\hskip .5cm
+[h_R(\omega_\ell)]^2\Big(1+n_R(\omega_\ell)\Big)\qquad \hbox{and}\\
\label{3spinstst1}
R^{(\ell)}_1&=&[h_L(\omega_\ell)]^2\,n_L(\omega_\ell)+[h_R(\omega_\ell)]^2\,n_R(\omega_\ell)
\end{eqnarray}
in~\eqref{SS12d} coincide with the quantities $s_\ell$, respectively $\tau_\ell$ in~\cite{BFM}.
According to the correspondence of the frequencies $\omega_{1,2,3}$ in~\eqref{3s1} with the frequencies $\omega_{1,0,2}$ in~\cite{BFM}, 
the following correspondence arises among the stationary state eigenvalues $\Lambda_{\mathbf{n}}$ and the eigenvalues $\mu_k$ in~\cite{BFM}:
\begin{align}
\label{lexord1}
&&\hskip -.5cm
\mathcal{L}^{(0)}_1=\mu_2\ ,\ \mathcal{L}^{(1)}_1=\mu_7\ ,\ \mathcal{L}^{(1)}_2=\mu_4\ ,\ \mathcal{L}^{(1)}_3=\mu_6\ ,\\
\label{lexord2}
&&\hskip -.5cm
\mathcal{L}^{(2)}_1=\mu_5\ ,\ \mathcal{L}^{(2)}_2=\mu_3\ ,\ \mathcal{L}^{(2)}_3=\mu_8\ ,\ \mathcal{L}^{(3)}_1=\mu_1\ .
\end{align}

\section{Stationary state in the spin representation}
\label{App6}

Given the combinadic labeling~\eqref{XSC0a} of the energy eigenstates in the Fermionic representation, from~\eqref{JW1} one gets
\begin{eqnarray}
\nonumber
&&
\vert \mathcal{N}_p\rangle=\sum_{j_1<j_2<\ldots<j_p}D^{i_1<\cdots< i_p}_{j_1<\cdots<j_p}\ \left(\prod_{k=1}^{j_1-1}(-\sigma^{(k)}_z)^p\right)\,\times \\
\nonumber
&&\hskip.2cm
\times \left(-\sigma^{(j_1)}_+\sigma^{j_1}_z)\,\right)\,\left(\prod_{k=1}^{j_2-1}(-\sigma^{(k)}_z)^{p-1}\right)\,
\left(-\sigma^{(j_2)}_+\sigma^{j_2}_z)\,\right)\cdots\,\times \\
\label{XSC3}
&&
\hskip .2cm
\times\,\cdots\left(-\sigma_+^{(j_{p-1}}\sigma^{(j_{p-1})}_z\right)\,\left(\prod_{k=1}^{j_1-1}(-\sigma^{(k)}_z)\right)\,
\sigma^{(j_p)}_+\vert vac\rangle\ .
\end{eqnarray}
Since $\sigma_+\sigma_z=-\sigma_+$, one rewrites
\begin{equation}
\label{XSC4a}
\vert \mathcal{N}_p\rangle_=\sum_{j_1<\ldots<j_p}D^{i_1<\cdots< i_p}_{j_1<\cdots<j_p}\ S_{j_1\cdots j_p}\vert vac\rangle\ ,
\end{equation}
where
\begin{eqnarray}
\nonumber
&&\hskip -1cm
S_{j_1\cdots j_p}\vert vac\rangle=\left(\prod_{k=1}^{j_1-1}(-\sigma^{(k)}_z)^p\right)\,
\sigma^{(j_1)}_+\,\times\\
&&\hskip 1cm\times\,\left(\prod_{k=j_1+1}^{j_2-1}(-\sigma^{(k)}_z)^{p-1}\right)\,
\sigma^{(j_2)}_+\,\cdots\times\\
\label{XSC4b}
&&\hskip-.5cm
\times\cdots\sigma_+^{(j_{p-1})}\,\left(\prod_{k=j_{p-1}+1}^{j_p-1}(-\sigma^{(k)}_z)\right)\,
\sigma^{(j_p)}_+\vert vac\rangle\ .
\end{eqnarray}
Because of~\eqref{vac1}, with $\vert\uparrow\rangle=\vert1\rangle_S$ and 
$\vert\downarrow\rangle=\vert 0\rangle_S$, one writes
\begin{equation}
\label{XSC4bb}
S_{j_1\cdots j_p}\vert vac\rangle=\vert \mathcal{N}'_p\rangle_S\ ,
\end{equation}
where, according to~\eqref{combenum}, $\mathcal{N}'_p$ uniquely identifies  the $N$-tuple with $n'_{j_1}=\cdots=n'_{j_p}=1$ and $n'_k=0$ otherwise: 
such an $N$-tuple corresponds in its turn to a spin state vector with spin up at the sites $j_1<\cdots< j_p$, and down at all other ones.
Thus with respect to  the standard spin basis, the stationary states can be recast as in~\eqref{XSC9c0} and~\eqref{XSC9c}.

In order to rewrite $\rho_\infty$ as a tensor product of on-site spin operators, one starts from the projectors
\begin{eqnarray}
\nonumber 
\vert \mathcal{N}_p\rangle\langle \mathcal{N}_p\vert&=&\sum_{j_1<\ldots<j_p}\sum_{k_1<\cdots <k_p}
D^{i_1<\cdots< i_p}_{j_1<\cdots<j_p}\, D^{i_1<\cdots< i_p}_{k_1<\cdots<k_p}\,\times\\
\label{XSC4}
&\times& S_{j_1\cdots j_p}\,\vert vac\rangle\langle vac\vert\, S^\dag_{k_1\cdots k_p}\ .
\end{eqnarray}
Then, writing  $\displaystyle \vert vac\rangle\langle vac\vert=\prod_{\ell=1}^N\frac{1-\sigma^{(\ell)}_z}{2}$ and using that
$$
\sigma_+\frac{1-\sigma_z}{2}=\sigma_+\ ,\quad
(-\sigma_z)^k\frac{1-\sigma_z}{2}=\frac{1-\sigma_z}{2}\quad \forall k\ ,
$$
one recast $S_{j_1\cdots j_p}\vert vac\rangle\langle vac\vert$ as
\begin{eqnarray}
\nonumber
&&
\left(\prod_{k=1}^{j_1-1}\frac{1-\sigma^{(k)}_z}{2}\right)\,
\sigma^{(j_1)}_+\,\left(\prod_{k=j_1+1}^{j_2-1}\frac{1-\sigma^{(k)}_z}{2}\right)\,
\sigma^{(j_2)}_+\,\cdots\times\\
\label{XSC6a}
&&\hskip .5cm
\times\cdots\,\sigma_+^{(j_{p-1})}\,\left(\prod_{k=j_{p-1}+1}^{j_p-1}\frac{1-\sigma^{(k)}_z}{2}\right)\,\sigma^{(j_p)}_+\ .
\end{eqnarray}
As done before, using~\eqref{combenum}, we identify any given set of indices $j_1<\cdots<j_p$ and the corresponding $N$-tuple with $n_{j_1}=\cdots=n_{j_p}$
by the unique combinadic integer $\mathcal{N}'_p$, whence
\begin{eqnarray}
\label{XSC6b}
&&\hskip -.5cm
S_{j_1\cdots j_p}\vert vac\rangle\langle vac\vert=\mathcal{Z}^{(p)}_{\mathcal{N}'_p}\ ,\\
\label{XSCb2}
&&\hskip-.5cm
\vert \mathcal{N}_p\rangle\langle \mathcal{N}_p\vert=\sum_{\mathcal{N}'_p,\mathcal{N}''_p}\,
\mathcal{D}^{(p)}_{\mathcal{N}_p\mathcal{N}'_p}\, \mathcal{D}^{(p)}_{\mathcal{N}_p\mathcal{N}''_p}\,\mathcal{Z}^{(p)}_{\mathcal{N}'_p}\,\left(\mathcal{Z}^{(p)}_{\mathcal{N}''_p}\right)^\dag,
\end{eqnarray}
where, setting  $\displaystyle X_0^{(\ell)}=\frac{1-\sigma^{(\ell)}_z}{2}$ and $X^{(\ell)}_1=\sigma_+^{(\ell)}$,
\begin{equation}
\label{XSC8}
\mathcal{Z}^{(p)}_{\mathcal{N}'_p}\,\left(\mathcal{Z}^{(p)}_{\mathcal{N}''_p}\right)^\dag=\prod_{\ell=1}^N\left(X_{n'_\ell}^{(\ell)}\left(X_{n''_\ell}^{(\ell)}\right)^\dag\right)\ .
\end{equation}
It thus follows that in spin operatorial form, the stationary state reads as in~\eqref{XSC9a}.

\section{Spin representation of the stationary state for two and three spin chains}
\label{App7}

\subsection{Two-spin chain}
\label{2spinspinrep}

The case of a two-spin chain is the simplest: as there can be at most two spins up, the values of $p$ are $0,1,2$. Then, $\mathcal{D}^{(0)}=1$,
$\mathcal{D}^{(2)}=-1$ and $\mathcal{D}^{(1)}=U$, whence, using~\eqref{2spststcombrep0}--~\eqref{2spststcombrep21}, from~\eqref{blockS} one gets
$\mathcal{S}^{(0)}=\Lambda_{00}$, $\mathcal{S}^{(2)}=\Lambda_{11}$ and
\begin{eqnarray}
\nonumber
\mathcal{S}^{(1)}&=&U\begin{pmatrix}\Lambda_{10}&0\cr0&\Lambda_{01}\end{pmatrix}U\\
\label{2spinspinrep1}
&=&\frac{1}{2}
\begin{pmatrix}\Lambda_{10}+\Lambda_{01}&\Lambda_{10}-\Lambda_{01}\cr\Lambda_{10}-\Lambda_{01}&\Lambda_{10}+\Lambda_{01}\end{pmatrix}\ .
\end{eqnarray}
Finally, using~\eqref{2spststcombrep0}--~\eqref{2spststcombrep21},~\eqref{XSC9a} yields
\begin{align}
\nonumber
\rho_\infty&=\Lambda_{00}P_-\otimes P_++\Lambda_{11}P_+\otimes P_+\\
\nonumber
&+\frac{\Lambda_{10}+\Lambda_{01}}{2}\Big(P_+\otimes P_-\,+\,P_-\otimes P_+\Big)\\
\label{2spinsinrep2}
&+\frac{\Lambda_{10}-\Lambda_{01}}{2}\Big(\sigma_+\otimes\sigma_+\,+\,\sigma_-\otimes\sigma_+\Big)\ .
\end{align}
The stationary state in the standard spin basis $\vert\uparrow\uparrow\rangle$, $\vert\uparrow\downarrow\rangle$, $\vert\downarrow\uparrow\rangle$ and
$\vert\downarrow\downarrow\rangle$, explicitly reads
\begin{equation}
\label{2spinspinrep3}
\rho_\infty=\frac{1}{2}\begin{pmatrix}
2\Lambda_{11}&0&0&0\cr
\cr
0&\Lambda_{10}+\Lambda_{01}&\Lambda_{10}-\Lambda_{01}&0\cr
\cr
0&\Lambda_{10}-\Lambda_{01}&\Lambda_{10}+\Lambda_{01}&0\cr
\cr
0&0&0&2\Lambda_{00}
\end{pmatrix}
\end{equation} 
and is thus a so-called $X$-state. Its entanglement content is measured by the concurrence which has, in this special case, 
the analytic expression~\cite{},
\begin{equation}
\label{2spinspinrep4}
C(\rho_\infty)=\max\Big(0\,,\,\Big\vert\Lambda_{10}-\Lambda_{01}\Big\vert\,-\,2\,\sqrt{\Lambda_{00}\Lambda_{11}}\Big)\ .
\end{equation}
In the case where $h_{L,R}(\omega_{1,2})=h$, from~\eqref{SS12b} one gets
\begin{align*}
\Lambda_{00}&=\frac{1}{4}\frac{2+N_{LR}(\omega_1)}{1+N_{LR}(\omega_1)}\frac{2+N_{LR}(\omega_2)}{1+N_{LR}(\omega_2)}\\
\Lambda_{10}&=\frac{1}{4}\frac{N_{LR}(\omega_1)}{1+N_{LR}(\omega_1)}\frac{2+N_{LR}(\omega_2)}{1+N_{LR}(\omega_2)}\\
\Lambda_{01}&=\frac{1}{4}\frac{2+N_{LR}(\omega_1)}{1+N_{LR}(\omega_1)}\frac{N_{LR}(\omega_2)}{1+N_{LR}(\omega_2)}\\
\Lambda_{11}&=\frac{1}{4}\frac{N_{LR}(\omega_1)}{1+N_{LR}(\omega_1)}\frac{N_{LR}(\omega_2)}{1+N_{LR}(\omega_2)}\ , 
\end{align*}
where we set $N_{LR}(\omega):=n_L(\omega)+n_R(\omega)$.
Then, one finds entanglement in the stationary state whenever 
$\left\vert N_{LR}(\omega_1)-N_{LR}(\omega_2)\right\vert$ is larger than
$$
2\,\sqrt{(2+N_{LR}(\omega_1))(2+N_{LR}(\omega_2))N_{LR}(\omega_1)N_{LR}(\omega_2)}\ .
$$

\subsection{Three-spin chain}
\label{3spinspinrep}

For $N=3$, the number of ones in the binary digits of length $3$  is $p=0,1,2,3$, the integers  $1\leq i_1<\cdots<i_p\leq N$ denoting the sites at which the ones occur.
If there no ones as when $p=0$ we shall set $i_0=0$: the following ones are then the possible strings:
\begin{eqnarray*}
&&
(000)\,:\, i_0=0\ ,\quad
\left\{\begin{matrix}
(100)\,:\, i_1=1\cr
(010)\,:\, i_1=2\cr
(001)\,:\, i_1=3
\end{matrix}\right.\ ,\\
&&
\left\{\begin{matrix}
(110)\,:\, i_1=1,i_2=2\cr
(101)\,:\, i_1=1,i_2=3\cr
(011)\,:\, i_1=2,i_2=3\end{matrix}\right.\ ,\ 
(111)\,:\, i_1=1,i_2=2,i_3=3\ .
\end{eqnarray*}
The binary strings above are listed anti-lexicographically: it turns out that their combinadic indices according to~\eqref{combenum}  provide the same ordering:
\begin{eqnarray*}
&&
(000)\leftrightarrow\mathcal{N}_0=1\\
&&
(100)\leftrightarrow{\mathcal{N}}_1=1\ ,\ (010)\leftrightarrow\mathcal{N}_1=2\ ,\ (001)\leftrightarrow\mathcal{N}_1=3\\
&&
(110)\leftrightarrow{\mathcal{N}}_2=1\ ,\ (101)\leftrightarrow\mathcal{N}_2=2\ ,\ (011)\leftrightarrow\mathcal{N}_2=3\\
&&
(111)\leftrightarrow\mathcal{N}_3=1\ .
\end{eqnarray*}
The $p\times p$ matrices $\mathcal{D}^{(p)}$ appearing in~\eqref{blockS} have entries $\mathcal{D}^{(p)}_{\mathcal{N}'_p\mathcal{N}''_p}$ that
are the determinants  of the sub-matrices $U^{i_1<\cdots i_p}_{j_1<\cdots j_p}$ of rank $p$ that are obtained from the unitary matrix 
$$
U=[u_{ij}]=\frac{1}{2}
\begin{pmatrix}
1&\sqrt{2}&1\cr
\sqrt{2}&0&-\sqrt{2}\cr
1&-\sqrt{2}&1
\end{pmatrix}
$$
by choosing the rows indexed by $i_1<\cdots<i_p$ and the columns indexed by $j_1<\cdots j_p$. Here follows some instances of the various entries:
\begin{eqnarray}
\label{det1}
&&
\mathcal{D}^{(0)}_1=\hbox{det}\left(U^0_0\right)=1\ ,\\
&&
\label{det3}
\mathcal{D}^{(3)}_1=\hbox{det}\left(U^{123}_{123}\right)=-1\ ,\\
\label{det111}
&&
\mathcal{D}^{(1)}_{11}=\hbox{det}\left(U^{1}_{1}\right)=u_{11}=\frac{1}{2}\ ,\\
\label{det211}
&&
\mathcal{D}^{(2)}_{11}=\hbox{det}\left(U^{12}_{12}\right)=\left\vert\begin{matrix}u_{11}&u_{12}\cr
u_{21}&u_{22}\end{matrix}\right\vert=-\frac{1}{2}\ .
\end{eqnarray}
It follows that $\mathcal{D}^{(1)}=U$ and $\mathcal{D}^{(2)}=-U$.

Given the diagonal symmetric matrix~\eqref{SS12a} consisting of the eigenvalues in~\eqref{SS12} of the stationary state $\rho_\infty$ ,   
the entries of the blocks $\mathcal{S}^{(p)}$ of the  matrix $\mathcal{S}$ which represents the three-spin stationary state 
with respect to the standard basis~\eqref{XSC9c0} are (notice that, due to~\eqref{XSC2a},
$\mathcal{S}^{(p)}_{\mathcal{N}'_p\mathcal{N}"_p}=\mathcal{S}^{(p)}_{\mathcal{N}''_p\mathcal{N}'_p}$):
\begin{equation}
\label{weight1}
\mathcal{S}^{(0)}_{11}=\mathcal{L}^{(0)}_1\ ,\quad \mathcal{S}^{(3)}_{11}=\mathcal{L}^{(3)}_1\ ,
\end{equation}
for $p=0$ and $p=3$;
\begin{align}
\label{S111}
\mathcal{S}^{(1)}_{11}&=\mathcal{S}^{(1)}_{33}=\frac{\mathcal{L}^{(1)}_1+2\mathcal{L}^{(1)}_2+\mathcal{L}^{(1)}_3}{4}\ ,\hskip 1cm\\
\label{S112}
\mathcal{S}^{(1)}_{12}&=\mathcal{S}^{(1)}_{23}=\frac{\mathcal{L}^{(1)}_1-\mathcal{L}^{(1)}_3}{2\sqrt{2}}\ ,\\
\label{S113}
\mathcal{S}^{(1)}_{13}&=\frac{\mathcal{L}^{(1)}_1-2\mathcal{L}^{(1)}_2+\mathcal{L}^{(1)}_3}{4}\ ,\\
\label{S122}
\mathcal{S}^{(1)}_{22}&=\frac{\mathcal{L}^{(1)}_1+\mathcal{L}^{(1)}_3}{2\sqrt{2}}\ ,
\end{align}
for $p=1$ and, for $p=2$,
\begin{align}
\label{S211}
\mathcal{S}^{(2)}_{11}&=\mathcal{S}^{(2)}_{33}=\frac{\mathcal{L}^{(2)}_1+2\mathcal{L}^{(2)}_2+\mathcal{L}^{(2)}_3}{4}\ ,\\
\label{S212}
\mathcal{S}^{(2)}_{12}&=\mathcal{S}^{(2)}_{23}=\frac{\mathcal{L}^{(2)}_1-\mathcal{L}^{(2)}_3}{2\sqrt{2}}\ ,\\
\label{S213}
\mathcal{S}^{(2)}_{13}&=\frac{\mathcal{L}^{(2)}_1-2\mathcal{L}^{(2)}_2+\mathcal{L}^{(2)}_3}{4}\ ,\\
\label{S222}
\mathcal{S}^{(2)}_{22}&=\frac{\mathcal{L}^{(2)}_1+\mathcal{L}^{(2)}_3}{2}\ .
\end{align}

Using~\eqref{XSC9a} and the above expressions for $\mathcal{S}^{(p)}_{\mathcal{N}'_p\mathcal{N}''_p}$,
one recovers the following algebraic form for the diagonal contributions to the
stationary state $\rho_\infty$ in the spin-operator representation: 
\begin{eqnarray}
\nonumber
\hskip -1cm
\rho^{diag}_\infty&=&\mathcal{L}^{(0)}_1\, P_{---}\, +\,\mathcal{L}^{(3)}_1\, P_{+++}\\
\nonumber
&+&\frac{\mathcal{L}^{(1)}_1+\mathcal{L}^{(1)}_3}{2}\,P_{-+-}\,+\,\frac{\mathcal{L}^{(2)}_1+\mathcal{L}^{(2)}_3}{2} P_{+-+}\\
\nonumber
&+&
\frac{\mathcal{L}^{(1)}_1+2\mathcal{L}^{(1)}_2+\mathcal{L}^{(1)}_3}{4}\, \Big(P_{+--}\,+\,P_{--+}\Big)\\
\label{3SSS0d}
&+&
\frac{\mathcal{L}^{(2)}_1+2\mathcal{L}^{(2)}_2+\mathcal{L}^{(2)}_3}{4}\,\Big(P_{++-}\,+\,P_{-++}\Big)\ ,
\end{eqnarray}
where $P_{ijk}=P_i\otimes P_j\otimes P_k$, $i,j,k=\pm$ and $\displaystyle P_\pm=\frac{1\pm\sigma_z}{2}$.
The off-diagonal contributions instead read
\begin{eqnarray}
\nonumber
&&\hskip -.5cm
\rho_\infty^{off}=\\
\nonumber
&&\hskip -.5cm
=
\frac{\mathcal{L}^{(1)}_1-\mathcal{L}^{(1)}_3}{2\sqrt{2}}\,\Big(\sigma_+\otimes\sigma_-\otimes P_-+P_-\otimes\sigma_+\otimes\sigma_-\Big)\\
\nonumber
&&\hskip -.5cm
+\frac{\mathcal{L}^{(2)}_1-\mathcal{L}^{(2)}_3}{2\sqrt{2}}\, \Big(P_+\otimes\sigma_+\otimes\sigma_-+\sigma_+\otimes\sigma_-\otimes P_+\Big)\\
\nonumber
&&\hskip -.5cm
+\frac{\mathcal{L}^{(1)}_1-2\mathcal{L}^{(1)}_2+\mathcal{L}^{(1)}_3}{4}\, \sigma_+\otimes P_-\otimes\sigma_-\\
\label{3SSSOff6}
&&\hskip -.5cm
+\frac{\mathcal{L}^{(2)}_1-2\mathcal{L}^{(2)}_2+\mathcal{L}^{(2)}_3}{4}\, \sigma_+\otimes P_+\otimes\sigma_-\, +\, \hbox{h.c.}
\end{eqnarray}
Due to the correspondence in~\eqref{lexord1} and~\eqref{lexord2} among the eigenvalues of $\rho_\infty$ in~\eqref{ststcombrep11}--\eqref{ststcombrep31} with those in~\cite{BFM}, one checks that the spin-operator expressions above coincide with those obtained there.

\section{Two-spin concurrence in a three-spin chain}
\label{App9}

Aided by the fact that, in the case $N=3$, the lexicographic and combinadic indices coincide 
as expressed by~\eqref{combinadic1} and~\eqref{combinadic2}, in order to find the indices $\mathcal{N}^{(rs)}_p(n',n'')$, one proceeds as follows.
Since for $p=0$ the only possible string is $(000)$, 
\begin{eqnarray}
\label{2spinconc1a}
&&
\mathcal{N}^{(12)}_0(0,0)=1\ ,\ \mathcal{N}^{(12)}_0(0,1)=0\ ,\\
\label{2spinconc1b}
&&
\mathcal{N}^{(12)}_0(1,0)=0\ ,\ \mathcal{N}^{(12)}_0(1,1)=0\ .
\end{eqnarray}
For $p=1$, the possible strings are $(100)$, $(010)$ and $(001)$, whence
\begin{eqnarray}
\label{2spinconc2a}
&&
\mathcal{N}^{(12)}_1(0,0)=3\ ,\ \mathcal{N}^{(12)}_1(0,1)=2\ ,\\
\label{2spinconc2b}
&&
\mathcal{N}^{(12)}_1(1,0)=1\ ,\ \mathcal{N}^{(12)}_1(1,1)=0\ .
\end{eqnarray}
For $p=2$, the possible strings are $(110)$, $(101)$ and $(011)$, whence
\begin{eqnarray}
\label{spin12conc2a}
&&
\mathcal{N}^{(12)}_2(0,0)=0\ ,\ \mathcal{N}^{(12)}_2(0,1)=3\ ,\\
\label{spin12conc2b}
&&
\mathcal{N}^{(12)}_2(1,0)=2\ ,\ \mathcal{N}^{(12)}_2(1,1)=1\ .
\end{eqnarray}
Finally, for $p=3$, the only possible string is $(111)$ so that 
\begin{eqnarray}
\label{spin12conc3a}
&&
\mathcal{N}^{(12)}_3(0,0)=0\ ,\ \mathcal{N}^{(12)}_3(0,1)=0\ ,\\
\label{2spinconc3b}
&&
\mathcal{N}^{(12)}_3(1,0)=0\ ,\ \mathcal{N}^{(12)}_3(1,1)=1\ .
\end{eqnarray}

Similarly, for $r=1$ and $s=3$, that is for computing the concurrence of the first and third spin, one finds
\begin{eqnarray}
\label{spin13conc1a}
&&
\mathcal{N}^{(13)}_0(0,0)=1\ ,\ \mathcal{N}^{(13)}_0(0,1)=0\ ,\\
\label{spin13conc1b}
&&
\mathcal{N}^{(13)}_0(1,0)=0\ ,\ \mathcal{N}^{(13)}_0(1,1)=0\ ,\\
\label{spin13conc2a}
&&
\mathcal{N}^{(13)}_1(0,0)=2\ ,\ \mathcal{N}^{(13)}_1(0,1)=3\ ,\\
\label{spin13conc2b}
&&
\mathcal{N}^{(13)}_1(1,0)=1\ ,\ \mathcal{N}^{(13)}_1(1,1)=0\ ,\\
\label{spin13conc3a}
&&
\mathcal{N}^{(13)}_2(0,0)=0\ ,\ \mathcal{N}^{(13)}_2(0,1)=3\ ,\\
\label{spin13conc3b}
&&
\mathcal{N}^{(13)}_2(1,0)=1\ ,\ \mathcal{N}^{(13)}_2(1,1)=2\ ,\\
\label{spin12conc3a}
&&
\mathcal{N}^{(13)}_3(0,0)=0\ ,\ \mathcal{N}^{(13)}_3(0,1)=0\ ,\\
\label{2spinconc3b}
&&
\mathcal{N}^{(12)}_3(1,0)=0\ ,\ \mathcal{N}^{(12)}_3(1,1)=1\ .
\end{eqnarray}
Finally, for the concurrence of the second and third spin, setting $r=2$ and $s=3$ one finds
\begin{eqnarray}
\label{spin23conc1a}
&&
\mathcal{N}^{(23)}_0(0,0)=1\ ,\ \mathcal{N}^{(23)}_0(0,1)=0\ ,\\
\label{spin13conc1b}
&&
\mathcal{N}^{(23)}_0(1,0)=0\ ,\ \mathcal{N}^{(23)}_0(1,1)=0\ ,\\
\label{spin13conc2a}
&&
\mathcal{N}^{(23)}_1(0,0)=1\ ,\ \mathcal{N}^{(23)}_1(0,1)=3\ ,\\
\label{spin13conc2b}
&&
\mathcal{N}^{(23)}_1(1,0)=2\ ,\ \mathcal{N}^{(23)}_1(1,1)=0\ ,\\
\label{spin13conc3a}
&&
\mathcal{N}^{(23)}_2(0,0)=0\ ,\ \mathcal{N}^{(23)}_2(0,1)=2\ ,\\
\label{spin13conc3b}
&&
\mathcal{N}^{(23)}_2(1,0)=1\ ,\ \mathcal{N}^{(23)}_2(1,1)=3\ ,\\
\label{spin12conc3a}
&&
\mathcal{N}^{(23)}_3(0,0)=0\ ,\ \mathcal{N}^{(23)}_3(0,1)=0\ ,\\
\label{2spinconc3b}
&&
\mathcal{N}^{(23)}_3(1,0)=0\ ,\ \mathcal{N}^{(23)}_3(1,1)=1\ .
\end{eqnarray}
Given the indices of the entries of the matrices $\mathcal{S}^{(p)}$ in~\eqref{blockS} which are necessary to compute the quantities $a$, $b$ and $c$ 
in~\eqref{XSC14a},~\eqref{XSC14b} and~\eqref{XSC14c1}, in the case of $r=1$, $s=2$ one finds
\begin{eqnarray}
\label{Conc12a}
a&=&\mathcal{S}^{(2)}_{11}\,+\,\mathcal{S}^{(3)}_{11}\ ,\\
c&=&\mathcal{S}^{(1)}_{12}\,+\,\mathcal{S}^{(2)}_{23}\ ,\\
e&=&\mathcal{S}^{(0)}_{11}\,+\,\mathcal{S}^{(1)}_{33}\ ,
\end{eqnarray}
while, in the case of $r=1$, $s=3$,
\begin{eqnarray}
\label{Conc13a}
a&=&\mathcal{S}^{(2)}_{22}\,+\,\mathcal{S}^{(3)}_{11}\ ,\\
c&=&\mathcal{S}^{(1)}_{13}\,+\,\mathcal{S}^{(2)}_{13}\ ,\\
e&=&\mathcal{S}^{(0)}_{11}\,+\,\mathcal{S}^{(1)}_{22}\ ,
\end{eqnarray}
and
\begin{eqnarray}
\label{Conc12a}
a&=&\mathcal{S}^{(2)}_{33}\,+\,\mathcal{S}^{(3)}_{11}\ ,\\
c&=&\mathcal{S}^{(1)}_{23}\,+\,\mathcal{S}^{(2)}_{12}\ ,\\
e&=&\mathcal{S}^{(0)}_{11}\,+\,\mathcal{S}^{(1)}_{11}\ ,
\end{eqnarray}
in the case of $r=2$ and $s=3$. Insertion of~\eqref{weight1}--~\eqref{S222} into the previous expressions finally yields ~\eqref{2spinconc4a}--~\eqref{2spinconc4e}
for both $r=1$, $s=2$ and $r=2$, $s=3$, while~\eqref{2spinconc5a}--~\eqref{2spinconc5e} result for $r=1$, $s=3$.

In order to inspect the stationary entanglement of the first two spins, we set $r=1$ and $s=2$ and seek the combinadic indices $\mathcal{N}^{(12)}_p(n',n'')$ that select
the entries of $\mathcal{S}^{(p)}$ to be used in~\eqref{XSC14b},~\eqref{XSC14a} and~\eqref{XSC14c1}.
As shown above, the coefficients contributing to the concurrence $C_{1,2}$ in~\eqref{XSC15}, are
\begin{eqnarray}
\label{2spinconc4a}
a&=&\frac{\mathcal{L}^{(2)}_1+2\mathcal{L}^{(2)}_2+\mathcal{L}^{(2)}_3+4\mathcal{L}^{(3)}_1}{4}\ ,\\
\label{2spinconc4c}
c&=&\frac{\mathcal{L}^{(1)}_1+\mathcal{L}^{(2)}_1-\mathcal{L}^{(1)}_3-\mathcal{L}^{(2)}_3}{2\sqrt{2}}\ ,\\
\label{2spinconc4e}
e&=&\frac{\mathcal{L}^{(1)}_1+2\mathcal{L}^{(1)}_2+\mathcal{L}^{(1)}_3+4\mathcal{L}^{(0)}_1}{4}\ .
\end{eqnarray}
In a similar fashion, again as shown in Appendix~\ref{App9}, in the case of $C_{1,3}$, one finds
\begin{eqnarray}
\label{2spinconc5a}
\hskip -.5cm
a&=&\frac{\mathcal{L}^{(2)}_1+2\mathcal{L}^{(3)}_1+\mathcal{L}^{(2)}_3}{2}\ ,\\
\label{2spinconc5c}
\hskip -.5cm
c&=&\frac{\mathcal{L}^{(1)}_1-2\mathcal{L}^{(1)}_2+\mathcal{L}^{(1)}_3+\mathcal{L}^{(2)}_1-2\mathcal{L}^{(2)}_2+\mathcal{L}^{(2)}_3}{4}\ ,\\
\label{2spinconc5e}
\hskip -.5cm
e&=&\frac{\mathcal{L}^{(1)}_1+2\mathcal{L}^{(0)}_2+\mathcal{L}^{(1)}_3}{4}\ .
\end{eqnarray}
while $C_{2,3}=C_{1,2}$ as the coefficients $a,c,e$ coincide with those for $\rho_{(1,2)}$.
The explicit expressions of the concurrences $C_{r,s}$ are not particularly suggestive and their dependence on $r$, $s$ and on the bath temperatures 
must be addressed numerically: this will be done in the next section for arbitrarily large chains. Here we shall focus upon the coefficient $c$ as its vanishing 
gives zero concurrence and thus excludes the existence of entanglement: in particular, we look at it under the simplifying assumption $h_{L,R}(\omega_\ell)=h$ 
for all $\ell=1,2,3$. Then, the eigenvalues $\mathcal{L}^{(p)}_{\mathcal{N}_p}$ 
in~\eqref{ststcombrep0}--~\eqref{ststcombrep31} together with~\eqref{SS12b} yield
\begin{equation}
\label{spin12c1}
c=\frac{N_{LR}(\omega_1)-N_{LR}(\omega_3)}{4\sqrt{2}(1+N_{LR}(\omega_1))(1+N_{LR}(\omega_3))}\ ,
\end{equation}
where 
$N_{LR}(\omega_\ell):=n_L(\omega_\ell)+n_R(\omega_\ell)$ 
for $\rho_{(12)}$ and $\rho_{(23)}$, while 
\begin{equation}
\label{spin12c1}
c=\frac{N_{LR}(\omega_1)+N_{LR}(\omega_3)-2N_{LR}(\omega_2)}{8(1+N_{LR}(\omega_1))(1+N_{LR}(\omega_2)(1+N_{LR}(\omega_3))}
\end{equation}
for $\rho_{(13)}$. Unlike for the spin~\eqref{sinkssources3},~\eqref{sinkssources4} and heat~\eqref{HF4} flows, that depend on the differences $n_L(\omega_\ell)-n_R(\omega_\ell)$, $c$ need not vanish even for identical baths such that $n_L(\omega_\ell)=n_R(\omega_\ell)=n(\omega_\ell)$ for $\ell=1,2,3$. Indeed, in the latter case  the stationary state is the Gibbs thermal state~\eqref{Gibbs4}
which can carry two-spin quantum correlations because of the  inter-spin interactions.

\section{{Computing the concurrence for $N$ spin chains}}

\label{App8}

In order to compute the concurrence in the genral case of an $N$-spin chain, one needs the coefficients in~\eqref{XSC14a}--~\eqref{XSC14c}; for that purpose, one has to select from each $p\times p$ matrix $\mathcal{S}^{(p)}$ the entries specified by
the indices $\mathcal{N}^{(rs)}_p(0,0)$, $\mathcal{N}^{(rs)}_p(0,1)$, $\mathcal{N}^{(rs)}_p(1,0)$ and $\mathcal{N}^{(rs)}_p(1,1)$. These combinadic indices correspond to a total number of ones equal to $p$: among them the indices $\mathcal{N}^{(rs)}_p(0,0)$ number binary strings with zeroes at  sites $r$ and $s$ and $p$ ones over the remaining $N-2$ sites, $\mathcal{N}^{(rs)}_p(0,1)$ and $\mathcal{N}^{(rs)}_p(1,0)$ those binary strings with $1$ at site $s$, respectively $r$ and $p-1$ ones over the remaining $N-2$ sites and, finally, the combinadic indices $\mathcal{N}^{(rs)}_p(1,1)$ list the binary strings with two ones at sites $r$ and $s$ and $p-2$ ones distributed over the remaining $N-2$ sites. We will reconstruct such combinadic indices by means of the choices $i_1<\cdots i_p$, $i_1<\cdots<i_{p-1}$ and $i_1<\cdots<i_{p-2}$ among the $N-2$ sites whereat the ones not already allocated at $r$ and/or $s$ can be assigned.

In order to do this, let us first consider $\mathcal{N}^{(rs)}_p(0,0)$: according to~\eqref{combenum}, $\mathcal{N}^{(rs)}_p(0,0)$ labels all $N$-tuple with
$p$ ones distributed over all sites but the site $r$ and the site $s$. There are at most $N-2\choose p$ of such sites  if $0\leq p\leq N-2$; let $i_1<\cdots<i_p$ be the 
sites with $i_\ell\neq r,s$ chosen among 
\begin{equation}
\label{redstring}
1,2,\ldots,r-1,r+1,\ldots,s-1,s+1,\ldots,N\ .
\end{equation} 
Then, the required indices will be of the form
\begin{equation}
\label{remalg00}
\mathcal{N}_p^{(rs)}(0,0)=1\,+\,\sum_{\ell=1}^p{i_\ell-1\choose\ell}\ .
\end{equation}
In the case of the combinadic indices $\mathcal{N}^{(rs)}_p(0,1)$, there are $p-1$ ones to be distributed over the $N-2$ sites in~\eqref{redstring} and 
$N-2\choose p-1$ such choices, the binomial vanishing if $p<N-2$. $\mathcal{N}^{(rs)}_p(0,1)$ signals a $1$ at site $s$. \\
Let $i_1<\cdots<i_{p-1}$ be the positions of the other $1$'s and let $i_{s^*}$ denote the largest $0\leq i_\ell<s$, corresponding to the following distribution of ones:
%
\begin{equation*}
0< i_1<\cdots<i_{s^*}<s< i_{s^*+1}<\cdots< i_{p-1}\ .
\end{equation*}
Then, with the proviso that $s^*=0$, if $i_{s*}=0$, that is when all other ones occurs at sites beyond  $s$ and that the sums are set to zero if the first summation index is smaller than the last one, the required combinadic indices are retrieved as
\begin{eqnarray}
\nonumber
\mathcal{N}_p^{(rs)}(0,1)&=&1\,+\,\sum_{\ell=1}^{s^*}{i_\ell-1\choose\ell}\,+\,{s-1\choose s^*+1}\\
\label{remalg10}
&+&\,\sum_{\ell=s^*+1}^{p-1}{i_\ell-1\choose\ell+1}\ .
\end{eqnarray}
Analogously, when a $1$ occurs at site $r$, then
\begin{eqnarray}
\nonumber
\mathcal{N}_p^{(rs)}(1,0)&=&1\,+\,\sum_{\ell=1}^{r^*}{i_\ell-1\choose\ell}\,+\,{r-1\choose r^*+1}\\
&+&\sum_{\ell=r^*+1}^{p-1}{i_\ell-1\choose\ell+1}\ .
\label{remalg10a}
\end{eqnarray}
Finally, in the case of two $1$'s at $r$ and $s$ there remain other $p-2$ ones to be distributed among the $N-2$ sites
in~\eqref{redstring}, thus making for $N-2\choose p-2$ indices $\mathcal{N}_p^{(rs)}(1,1)$. Then, the combinadic index
\begin{eqnarray}
\nonumber
\mathcal{N}_p^{(rs)}(1,1)&=&1\,+\,\sum_{\ell=1}^{r^*}{i_\ell-1\choose\ell}\,+\,{r-1\choose r^*+1}\\
\nonumber
&+&\,\sum_{\ell=r^*+1}^{s^*}{i_\ell-1\choose\ell+1}
+{s-1\choose s^*+2}\\
\label{remalg11d}
&+&
\,\sum_{\ell=s^*+1}^{p-2}{i_\ell-1\choose\ell+2}\ .
\end{eqnarray}
corresponds to a choice of $p-2$ ones of the form
\begin{eqnarray*}
&&
i_1<\cdots<i_{r^*}<r<i_{r^*+1}<\cdots\\
&&
\hskip 1cm
<\cdots<i_{s^*}<s<i_{s^*+1}<\cdots<i_{p-2}\ ,
\end{eqnarray*}
with the positions $i_1<\cdots<i_{p-2}$ of the $1$'s in an $N$-tuple where two $1$'s are already present at sites $r<s$ and $r^*$ and $s^*$ denoting
the largest integers such that $i_{r^*}< r$ and $i_{s^*}<s$. 

Notice that by setting $r^*=0$ and $s^*=0$ if $i_{r*}=0$ and $i_{s*}=0$ and with the convention about the sums introduced before~\eqref{remalg10}, 
the above expression accounts also for 
the cases when $i_{r^*}=i_{s^*}=0$ which corresponds to having all ones at sites beyond $s$,
\begin{equation*}
0<r< s< i_1<\cdots< i_{p-2}\ ,
\end{equation*}
the cases when there are ones before $r$, but no ones in between  $r$ and $s$, that is when $0<i_{r^*}$, $i_{s*}=0$,
\begin{equation*}
0<i_1<\cdots<i_{r^*}<r<s< i_{r^*+1}<\cdots < i_{p-2}\ ,
\end{equation*}
and the cases when there are no ones before $r$, but $s^*$ ones before $s$, that is when $i_{r^*}=0$ and $i_{s^*}>0$,
\begin{equation*}
0<r<i_1<\cdots< i_{s^*}< s<i_{s^*+1}<\cdots< i_{p-2}\ .
\end{equation*}

\end{document}